\documentclass[usenatbib]{mnras}
\usepackage{epsfig,booktabs,caption,times,graphicx,amsmath}

\def\Hy@FixNotFirstPage{%
	\gdef\Hy@FixNotFirstPage{%
		\setbox\AtBeginShipoutBox=\hbox{%
			\copy\AtBeginShipoutBox
		}%
	}%
}
\AtBeginShipout{\Hy@FixNotFirstPage}
 
\def\I{\,\textsc{i}}

\def\hst{{\it HST}}

\def\spitzer{{\it Spitzer}}

\title[Progenitor of SN~2017eaw]{The Dusty Progenitor Star of the Type II Supernova 2017eaw}

\author[Kilpatrick \& Foley]{Charles D. Kilpatrick$^1$\thanks{Email:
    cdkilpat@ucsc.edu} and Ryan~J.~Foley$^1$ \\
	$^1$Department of Astronomy and Astrophysics, University of California, Santa Cruz, CA 95064, USA}

\begin{document}
\date{Accepted 0000, Received 0000, in original form 0000}
\maketitle
\label{firstpage}

\begin{abstract}
\noindent

We present pre-explosion photometry of the likely progenitor star of the Type II supernova (SN~II) 2017eaw in NGC 6946.  We use a {\it Hubble Space Telescope} (\hst) image of SN~2017eaw to perform relative astrometry with \hst\ and {\it Spitzer Space Telescope} (\spitzer) imaging, finding a single point source consistent with its position.  We detect the progenitor star in $>$40 epochs of \hst\ and \spitzer\ imaging covering 12.9 years to 43 days before discovery.  While the progenitor luminosity was roughly constant for most of this period, there was a $\sim$20\% increase in its $4.5~\mu$m luminosity over the final 3 years before explosion.  We interpret the bright mid-infrared emission as a signature of circumstellar dust around the progenitor system. Using the pre-explosion photometry and assuming some circumstellar dust, we find the progenitor is most likely a red supergiant with $\log(L/L_{\odot}) = 4.9$ and $T = 3350$~K, obscured by a $>2\times10^{-5}~M_{\odot}$ dust shell with $R = 4000~R_{\odot}$ and $T = 960$~K.  Comparing to single-star evolutionary tracks, we find that the progenitor star had an initial mass of $13~M_{\odot}$ and a mass-loss rate of $2\times10^{-7}~M_{\odot}~\text{yr}^{-1}$, consistent with the population of SN~II progenitor stars.

\end{abstract}


\section{INTRODUCTION}\label{sec:introduction}

Stars with masses $>8~M_{\odot}$ undergo core collapse when their iron cores increase in mass and become unstable. The majority of these stars are the red supergiant (RSG) progenitor stars of Type~II supernovae (SNe~II).  This connection between SNe~II and RSGs is robustly predicted through comparison of observed supernova rates to the initial mass function \citep{smith+11} and measurements of the progenitor star radii from shock cooling models \citep{rubin+16}.  It is also directly shown through the growing sample of resolved RSG progenitor stars of SNe~II \citep[for a review see][]{smartt09}.

Detailed analysis of the population of RSG SN~II progenitor stars indicates that none of them have luminosities above $\log(L/L_{\odot})\approx5.2$ \citep[i.e., with initial masses above $17~M_{\odot}$;][]{smartt15}.  This observation is in conflict with the hypothesis that all RSGs result in SNe II and the observed luminosities of Galactic RSGs, which extend up to $\log(L/L_{\odot})=5.5$--$5.6$ \citep[i.e., with initial masses up to $25~M_{\odot}$;][]{levesque+05,massey+09}.  Assuming a Salpeter initial mass function, roughly 66\% of RSG stars with initial masses above $8.5~M_{\odot}$ ought to have $M_{init}>16.5~M_{\odot}$ \citep[see][for a detailed analysis]{smith+11}.  However, there are now over $13$ robust detections of RSG SN progenitor stars in the literature, none of which fall in this mass range \citep[e.g., SNe~2003gd, 2004A, 2004et, 2005cs, 2006my, 2008bk, 2009hd, 2009kr, 2009md, 2012A, 2012aw, 2012ec, 2016cok;][]{smartt+04,maund+09a,maund+09b,fraser+10,crockett+11,elias-rosa+11,fraser+11,maund+13,tomasella+13,fraser+14,maund+14,kochanek+17}. It is extremely unlikely that all of the observed RSG SN progenitors would have had initial masses $<17~M_{\odot}$ assuming they all come from a typical mass function.  This apparent conflict is the so-called ``red supergiant problem.''

One solution to this problem is that there is in fact a maximum mass for RSG SN progenitor stars above which RSGs do not produce SNe. \citet{smartt09} and \citet{smartt15} determine statistically that the current RSG SN progenitor star luminosity estimates are consistent with a maximal mass in the range of $16$--$21~M_{\odot}$.  Theoretical predictions indicate that some RSGs whose luminosities exceed this limit may instead produce ``failed SNe'' and direct collapse or fall back to a black hole \citep{woosley+12,lovegrove+13}.  These stars would disappear over a short timescale \citep{kochanek+08} or produce a low-luminosity, red transient with a weak shock breakout \citep{piro+13,lovegrove+13}.  \citet{gerke+15} report a potential example of a ``failed SN'' in NGC~6946.  The pre-explosion counterpart of this event was consistent with a $25~M_{\odot}$ RSG that increased gradually in luminosity over several hundred days and then promptly disappeared down to deep limits in optical bands \citep[although a $2000$--$3000~L_{\odot}$ infrared source remains;][]{adams+17}.

Even if some high-mass RSGs undergo prompt collapse to a black hole without a SN, it is likely that all SN~II progenitor stars are dust-obscured to some degree.  RSGs form dust in their winds \citep{verhoelst+09}, and analysis of resolved circumstellar environments around RSGs in the Milky Way indicates that these winds can form compact shells obscuring the underlying star.  SNe~II exhibit evidence for coronal line emission in early-time spectra \citep{khazov+15}, narrow, transient lines of hydrogen consistent with a compact shell of circumstellar material that is irradiated by the SN \citep{bullivant+18}, and excess mid-infrared emission consistent with heated dust in their circumstellar environments \citep{tinyanont+16}.  If high-mass RSGs are significantly dust-obscured in pre-SN imaging, it is possible that their luminosities and initial masses are underpredicted, or even that some of these stars go completely undetected.

\citet{walmswell+12} found that the current population of RSG SN progenitor stars could be consistent with a maximum initial mass of $21\substack{+2\\-1}~M_{\odot}$ assuming that all of them were obscured by dust from a RSG-like wind that was previously unaccounted for in analysis of their spectral energy distributions (SEDs).  Similarly, \citet{beasor+16} and \citet{davies+18} point out that pre-SN RSGs evolve to later spectral types, and so their luminosities are significantly underestimated when pre-explosion photometry is limited and matched to typical RSG SEDs. Analysis of the progenitor star of the SN~2012aw in \citet{kochanek+12} can account for absorption and scattering by circumstellar material and find a relatively low-mass progenitor star \citep[compared to, e.g.,][]{vandyk+12}, but analyses where the SED is so well-constrained are rare.  In addition, the total mass of circumstellar dust around SN~II progenitor systems is partly constrained by radio and X-ray observations \citep[e.g.,][]{chevalier+03,dwarkadas+12,dwarkadas+14}.  These studies suggest that there is not enough material to hide a high-mass RSG in some SN~II progenitor systems, but this finding emphasizes the role of circumstellar dust in shaping the observed SEDs of SN progenitor stars.

Here we discuss the SN~II 2017eaw in NGC~6946.  NGC~6946 is a well-studied galaxy and the host of over $10$ luminous transients in the past century \citep[e.g., SN~2002hh, 2004et, 2008S;][]{barlow+05,li+05,prieto+08}.  SN~2017eaw was discovered in NGC~6946 on 14.24 May 2017 by \citet{wiggins17}.  \citet{atel10374} spectroscopically identified SN~2017eaw as a SN~II on 14.75 May 2017, with broad H$\alpha$ and strong blue continuum emission.  Detailed photometric follow up confirmed that SN~2017eaw exhibited a plateau in its light curve \citep{tsvetkov+18}, consistent with the explosion of a star with an extended hydrogen envelope.  \citet{atel10378} identified a potential RSG progenitor star in pre-explosion \hst\ imaging, which was consistent with a RSG with $\log(L/L_{\odot})=4.9$ and an initial mass of $13~M_{\odot}$.  \citet{johnson+17} used $9$~yr of $UBV\!R$ imaging of the site of SN~2017eaw to demonstrate that its progenitor system was not extremely variable in optical bands.

We report detailed analysis of pre-explosion \hst\ and \spitzer\ imaging of the site of SN~2017eaw.  We identify a single point source consistent with being of the SN~2017eaw progenitor star.  This star is detected in multiple epochs of \hst/ACS and \spitzer/IRAC imaging where it exhibits a persistent mid-infrared excess consistent with predictions of a RSG surrounded by a compact circumstellar dust shell.  The source decreased by 30\% in the \hst/$F814W$ (roughly $I$) band over $12$~years and increased by 20\% in \spitzer\ $4.5~\mu$m emission around $1000$~days before discovery.  Following methods used in the analysis of the progenitor star of SN~2012aw by \citep{kochanek+12}, we fit a SED of a model RSG to optical to mid-infrared photometry of the SN~2017eaw counterpart from roughly $200$~days before core-collapse and determine that it was most likely a $13~M_{\odot}$ star surrounded by a relatively low mass dust shell.  Although the SN~2017eaw progenitor star is among the most massive known SN~II progenitor stars \citep[compared to examples in][]{smartt15}, it is consistent with predictions that the RSG progenitor stars of SN~II have an upper limit in mass.

Throughout this paper, we assume a distance to NGC~6946 of 6.72$\pm$0.15~Mpc \citep[derived from the tip of the red giant branch (TRGB) by][]{Tikhonov14}.  This value differs somewhat from $\approx$5.6--5.8~Mpc derived using SNe~II and the Tully-Fisher relation \citep[e.g., see][]{terry+02,sahu+06,rodriguez+14}, but the TRGB method is well-calibrated and this distance to NGC~6946 has already been used in the literature \citep[e.g.,][]{murphy+18,williams+18}.  For the Milky Way extinction to NGC~6946, we take $E(B-V)=0.30$~mag from \citet{schlafly+11}.

\section{DATA}\label{sec:observations}

\subsection{{\it Hubble Space Telescope}}\label{sec:hst}

We obtained \hst/ACS and WFC3 imaging of NGC~6946 from the Mikulski Archive for Space Telescopes\footnote{\url{https://archive.stsci.edu/}}.  These data were taken between 29 Jul. 2004 and 26 Oct. 2016 (SNAP-9788, PI Ho; GO-14156, PI Leroy; GO-14638, PI Long; GO-14786, PI Williams). The individual {\tt flc} and {\tt flt} files were processed using the relevant calibration files, including corrections for bias, dark current, flat-fielding, and bad-pixel masking. We combined the individual files from each epoch with {\tt DrizzlePac}, which performs cosmic-ray removal and image combination using the {\tt Drizzle} algorithm. With the drizzled images as a reference, we performed photometry on the individual {\tt flc/flt} frames using {\tt dolphot}\footnote{\url{http://americano.dolphinsim.com/dolphot/}}.  We used standard {\tt dolphot} parameters recommended for ACS/WFC and WFC3/IR.  The instrumental magnitudes were calibrated using zeropoints from the ACS/WFC zeropoint calculator tool for 29 Jul. 2004 and 26 Oct. 2016\footnote{\url{https://acszeropoints.stsci.edu/}} and using the WFC3/IR photometric zeropoints available at \url{http://www.stsci.edu/hst/wfc3/analysis/}.

\begin{figure*}
	\setlength{\fboxsep}{-1pt}
	\setlength{\fboxrule}{2pt}
	\begin{center}\begin{minipage}{6.8in}
	\fbox{\includegraphics[width=0.32\textwidth]{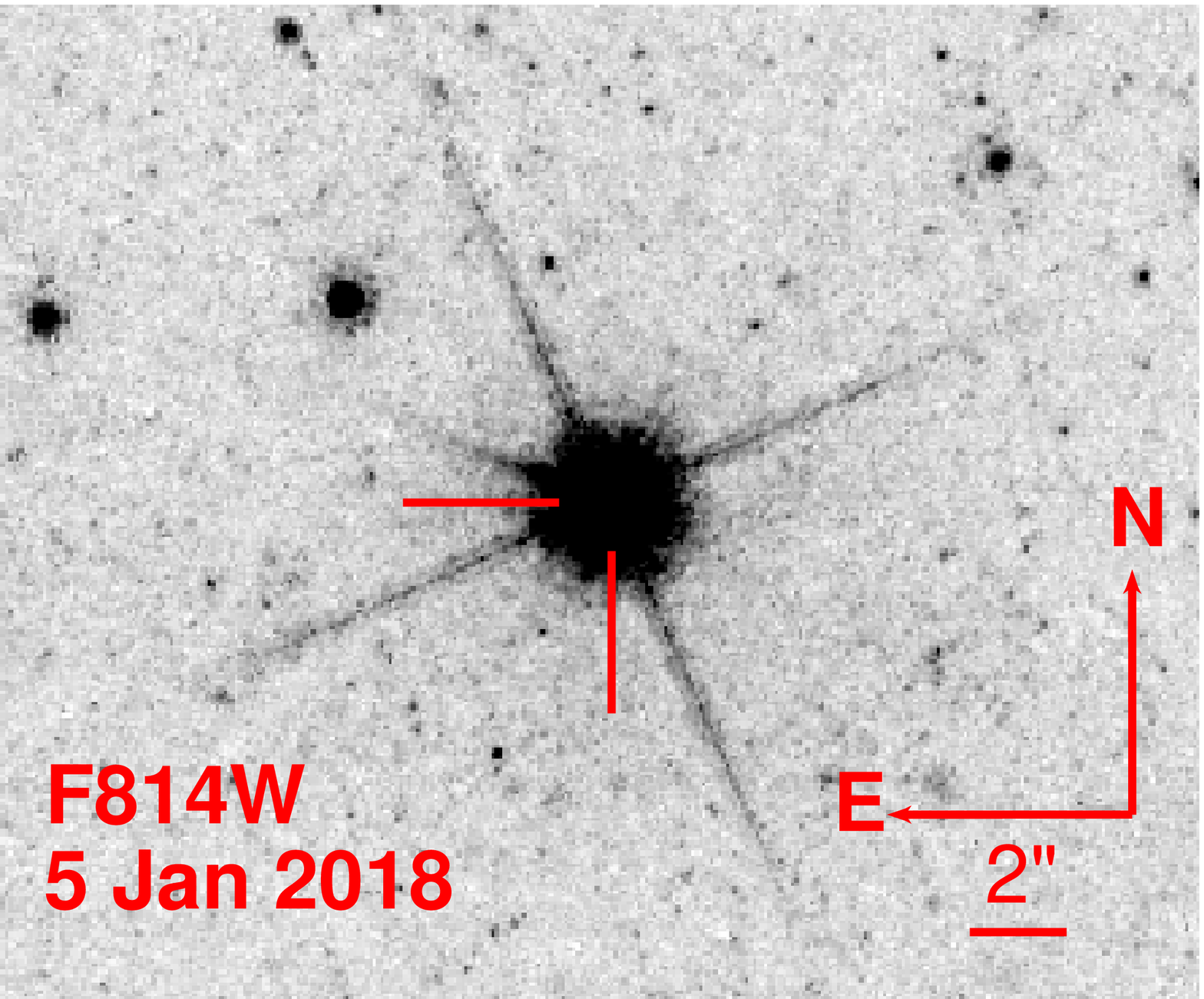}}
	\fbox{\includegraphics[width=0.32\textwidth]{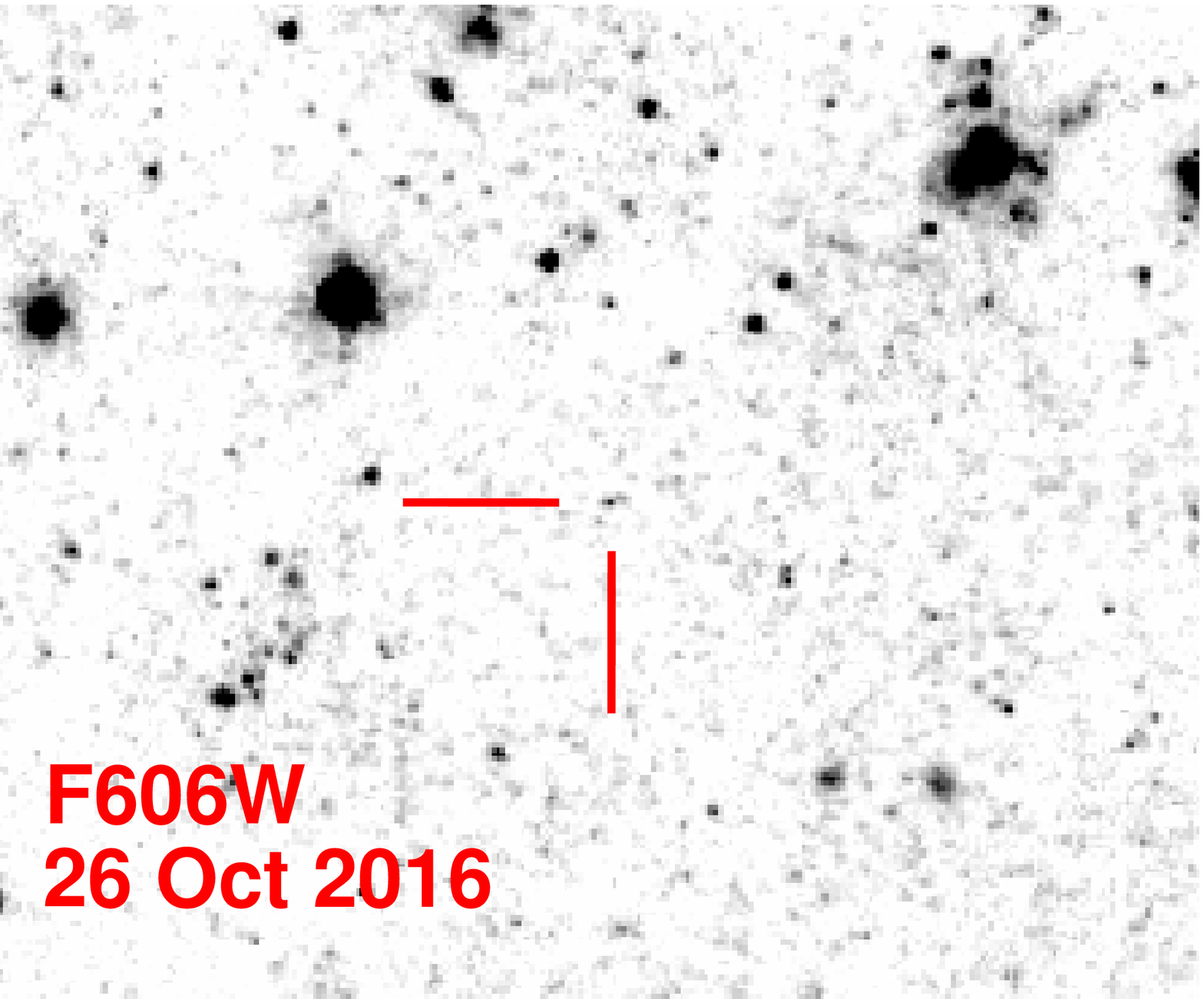}}
	\fbox{\includegraphics[width=0.32\textwidth]{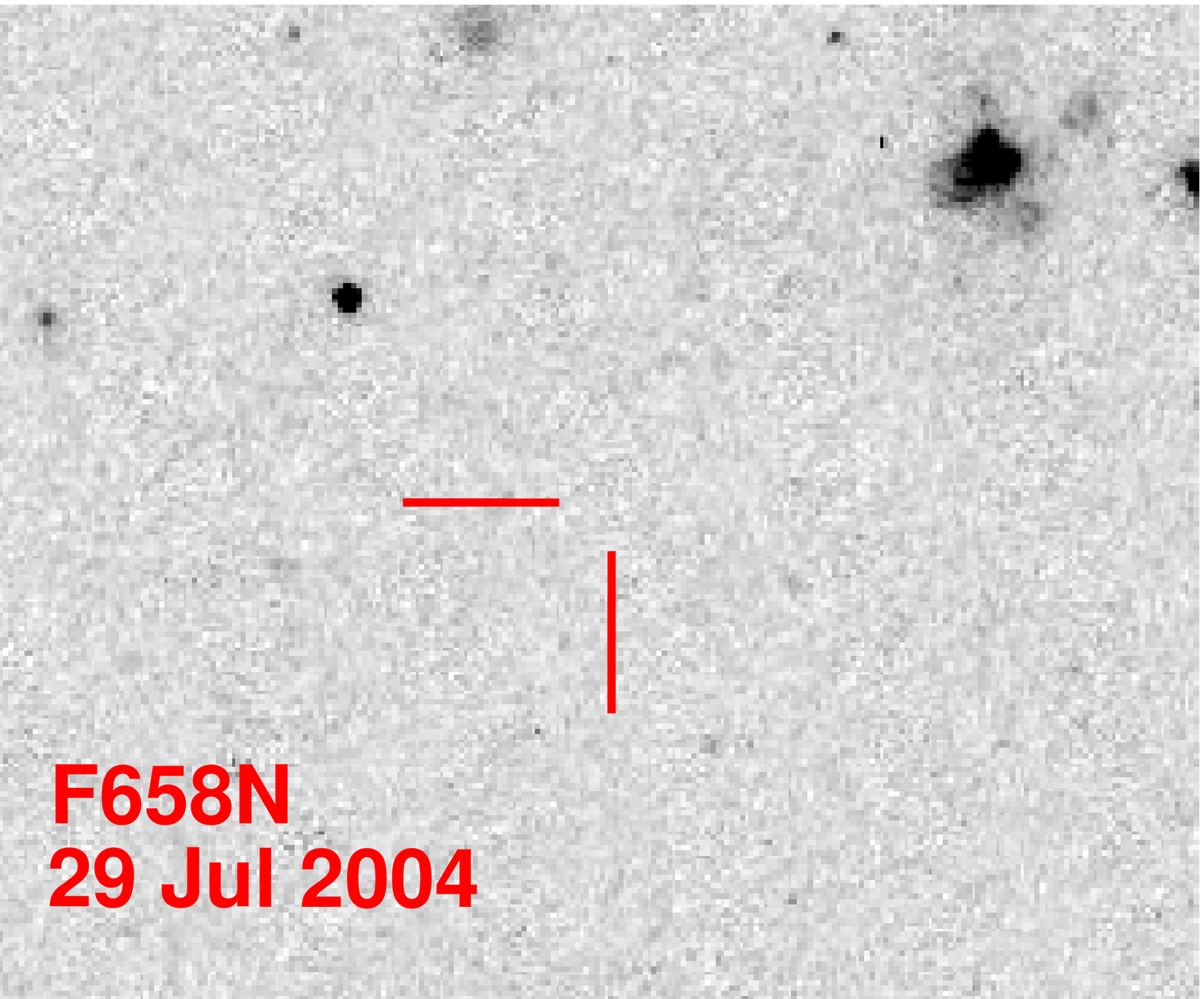}}
	\end{minipage}
	\begin{minipage}{6.8in}
	\fbox{\includegraphics[width=0.32\textwidth]{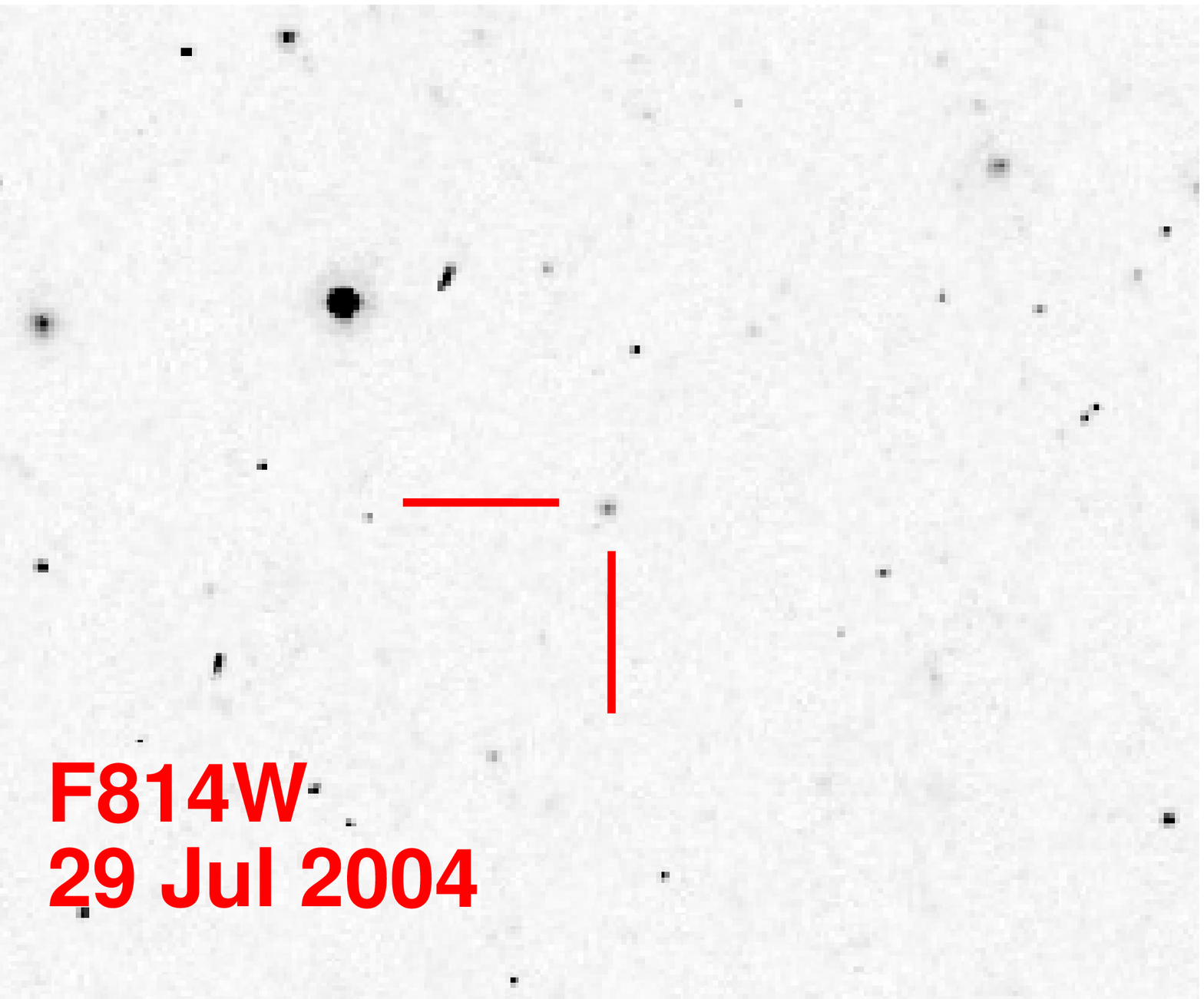}}
	\fbox{\includegraphics[width=0.32\textwidth]{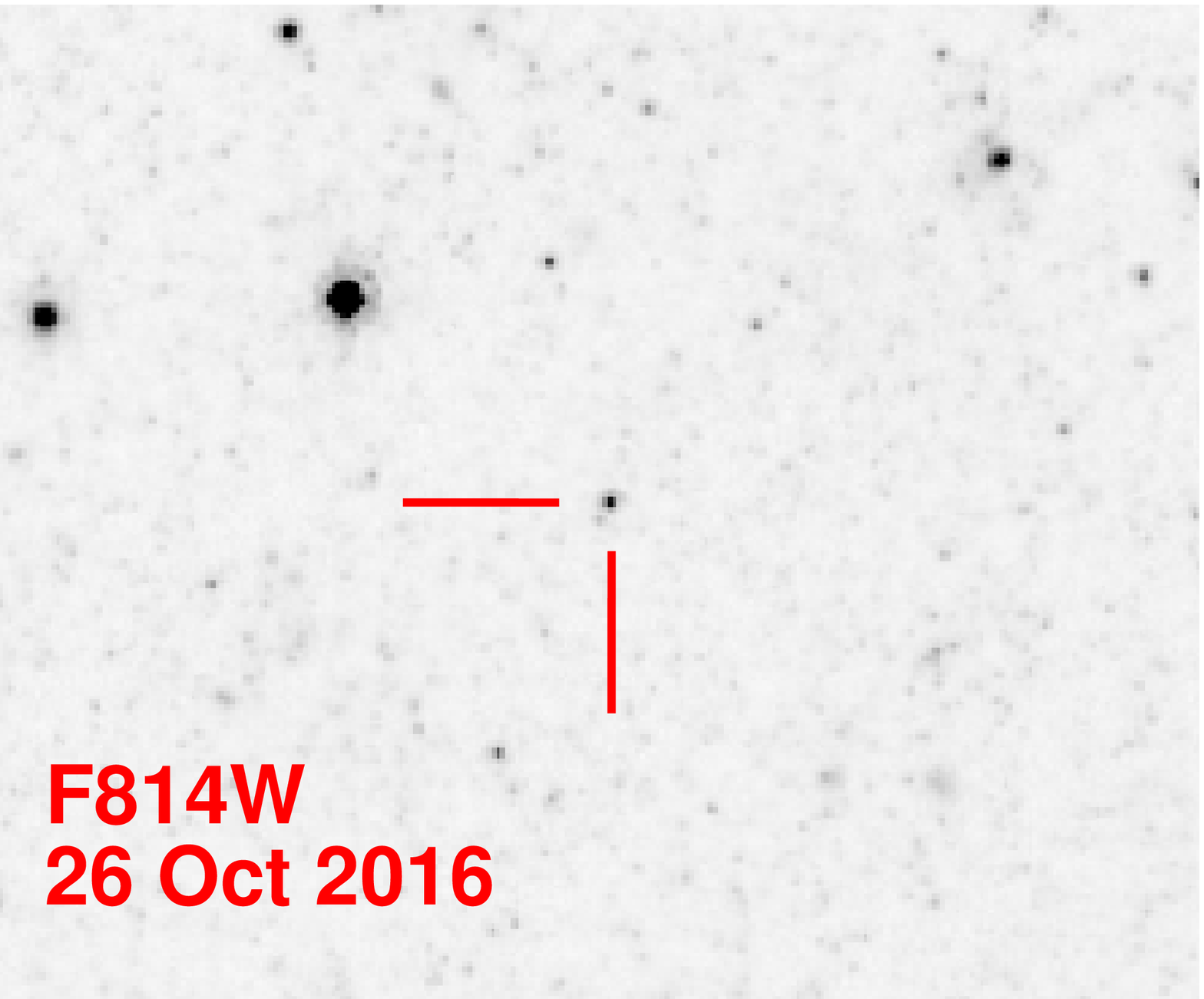}}
	\fbox{\includegraphics[width=0.32\textwidth]{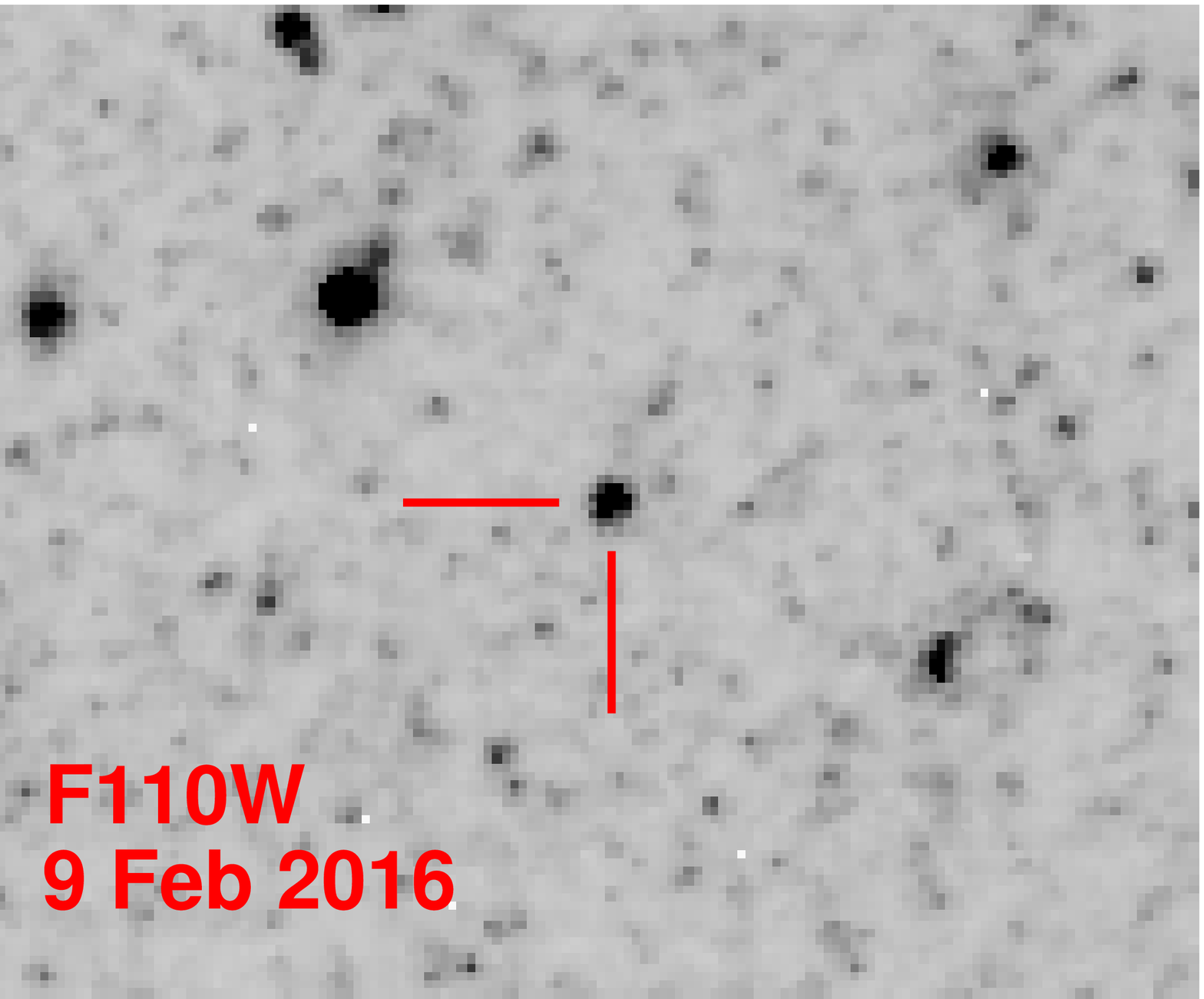}}
	\end{minipage}
	\begin{minipage}{6.8in}
	\fbox{\includegraphics[width=0.32\textwidth]{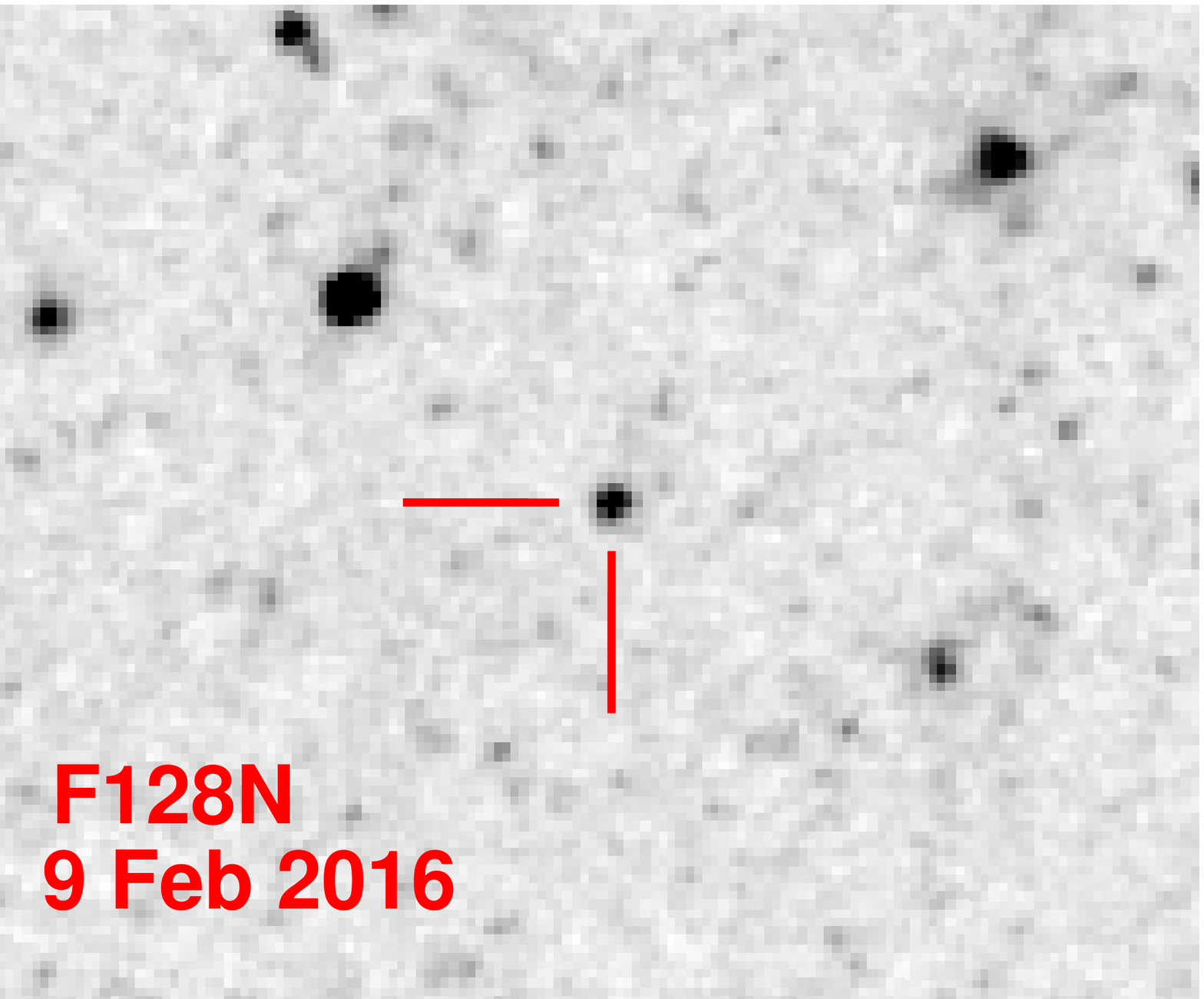}}
	\fbox{\includegraphics[width=0.32\textwidth]{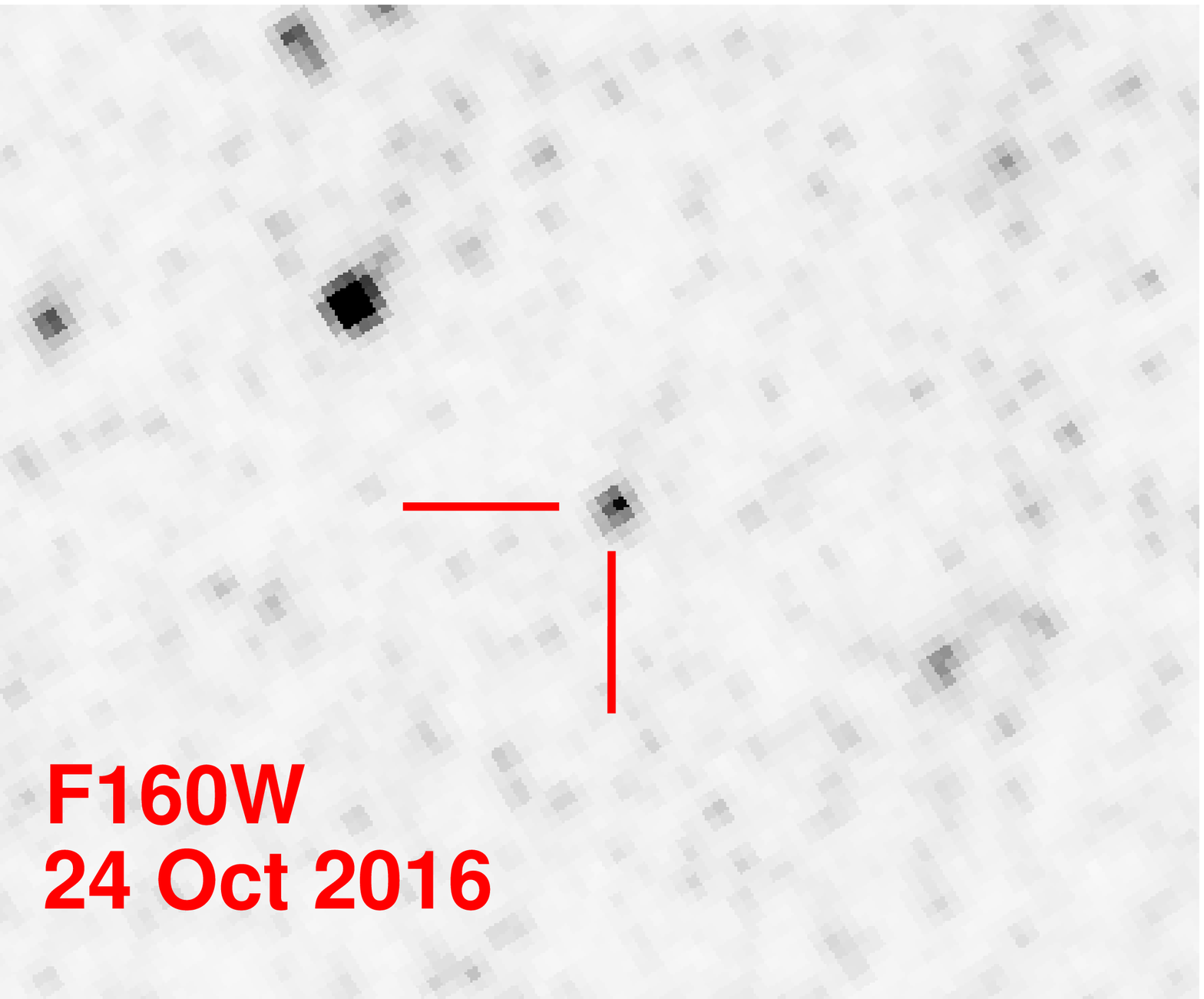}}
	\fbox{\includegraphics[width=0.32\textwidth]{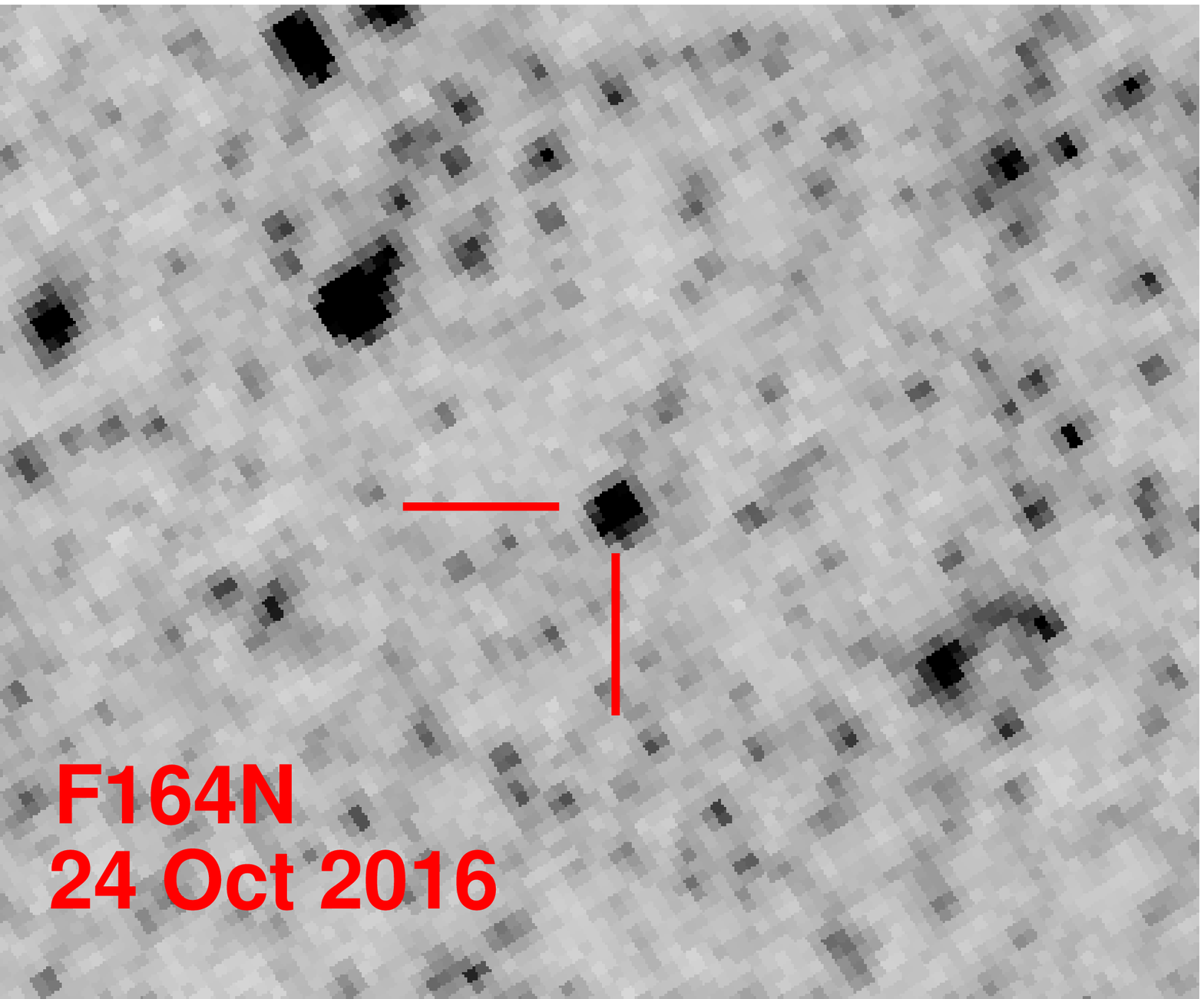}}
	\end{minipage}
	\end{center}
	\setlength{\fboxrule}{0.5pt}
	\caption{\hst/ACS and WFC3 imaging of a 12.3\arcsec $\times$ 10.4\arcsec\ region centered on SN~2017eaw (upper left) and the pre-explosion source.  The other panels show pre-explosion imaging, with the bandpass and observation date for each image is given in the lower-left the panel.  The locations of SN~2017eaw and its pre-explosion counterpart are denoted with red lines in each panel.}\label{fig:hst}
\end{figure*}

In addition, we obtained a single epoch of \hst/WFC3 $F814W$ imaging of SN~2017eaw obtained on 5 Jan. 2018 (SNAP-15166, PI Filippenko). These images consisted of $2\times390$~s frames, which we drizzled together using the same process described above and then performed photometry using {\tt dolphot}.  SN~2017eaw was easily identified in these images (\autoref{fig:hst}) and had a $F814W$ Vega magnitude of 15.290$\pm$0.004~mag.

\subsection{{\it Spitzer}}\label{sec:spitzer}

\begin{figure}
		\setlength{\fboxsep}{-1pt}
		\setlength{\fboxrule}{2pt}
		\fbox{\includegraphics[width=0.475\textwidth]{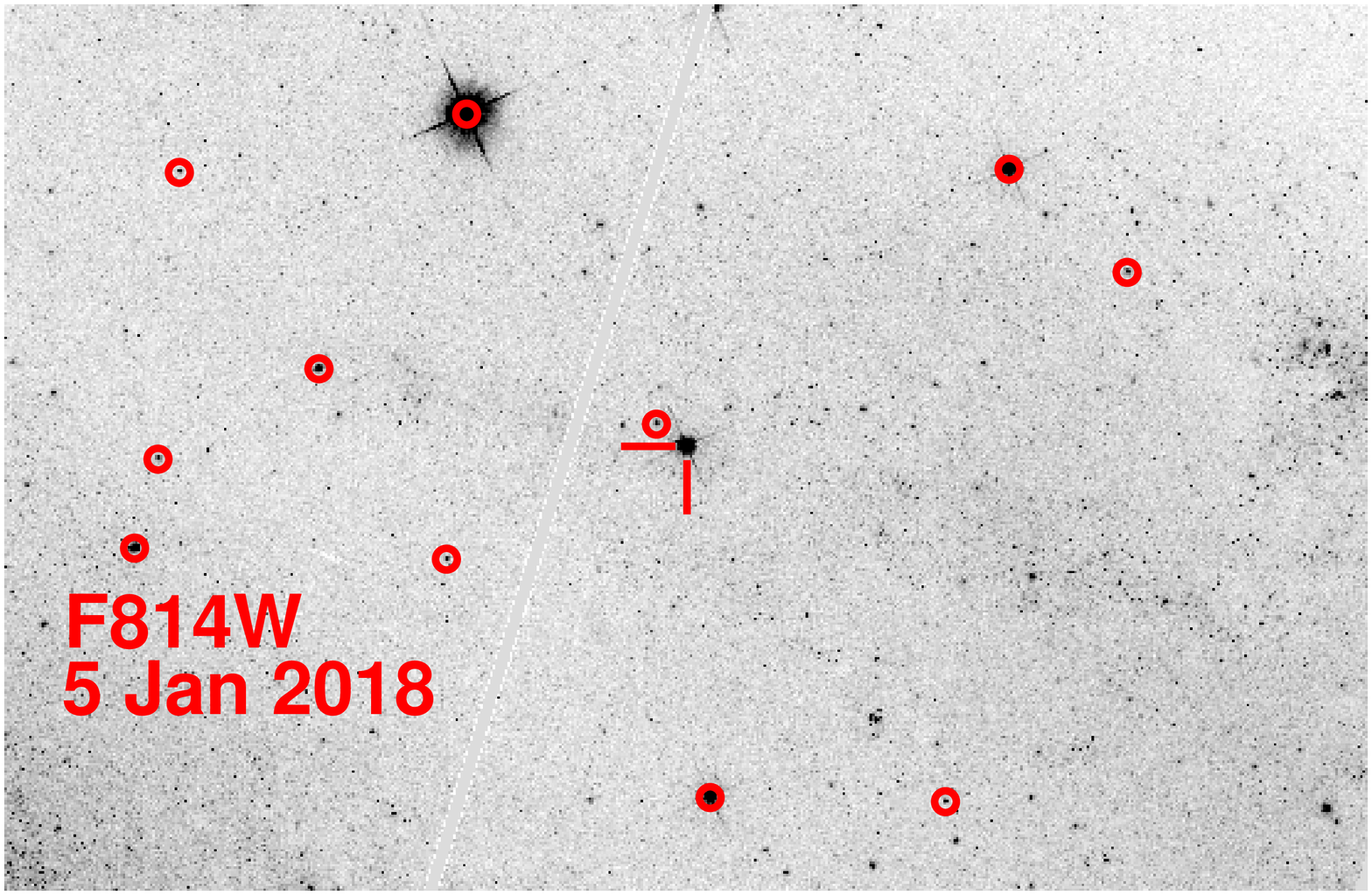}}
		\fbox{\includegraphics[width=0.475\textwidth]{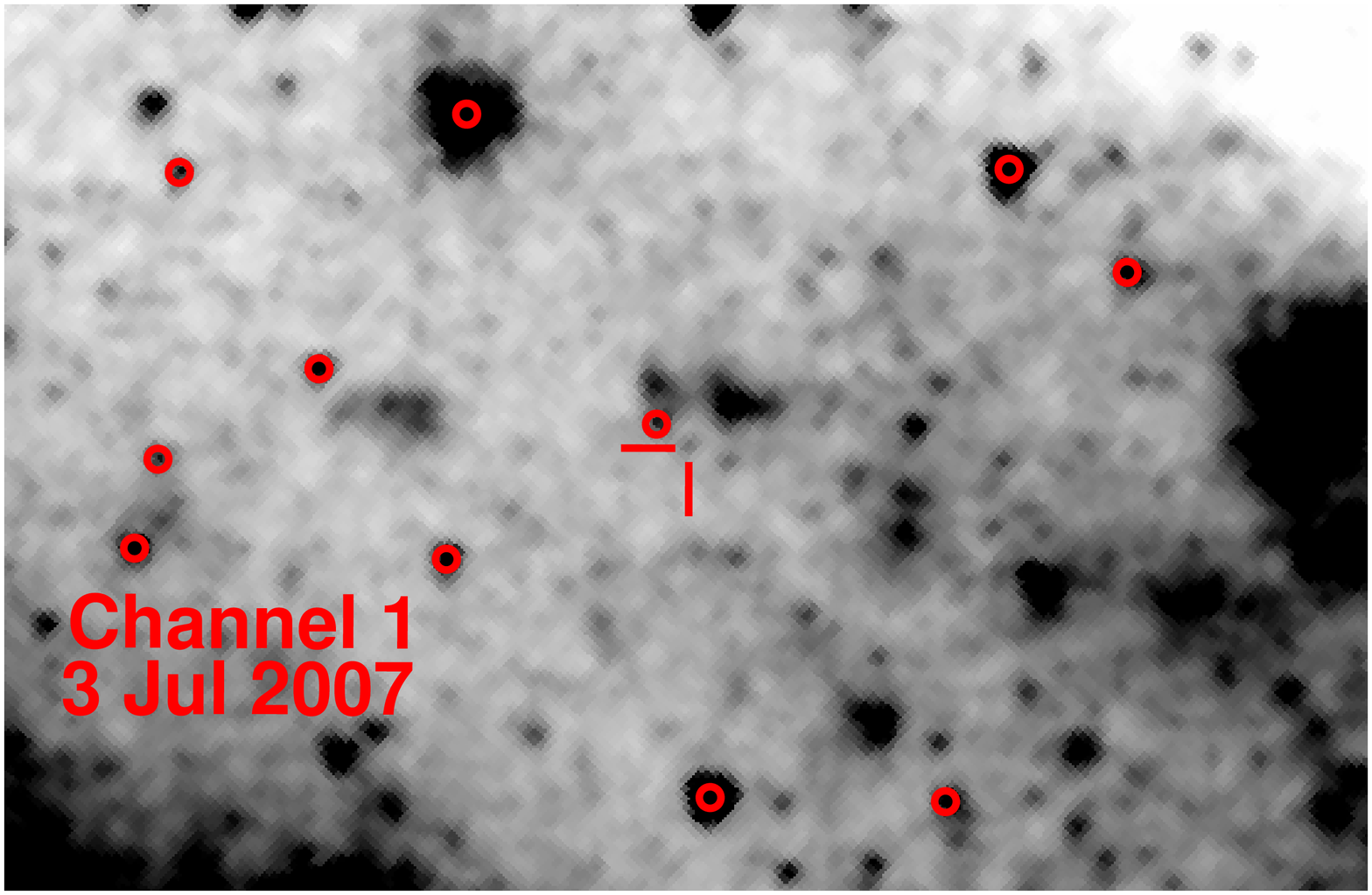}}
		\fbox{\includegraphics[width=0.475\textwidth]{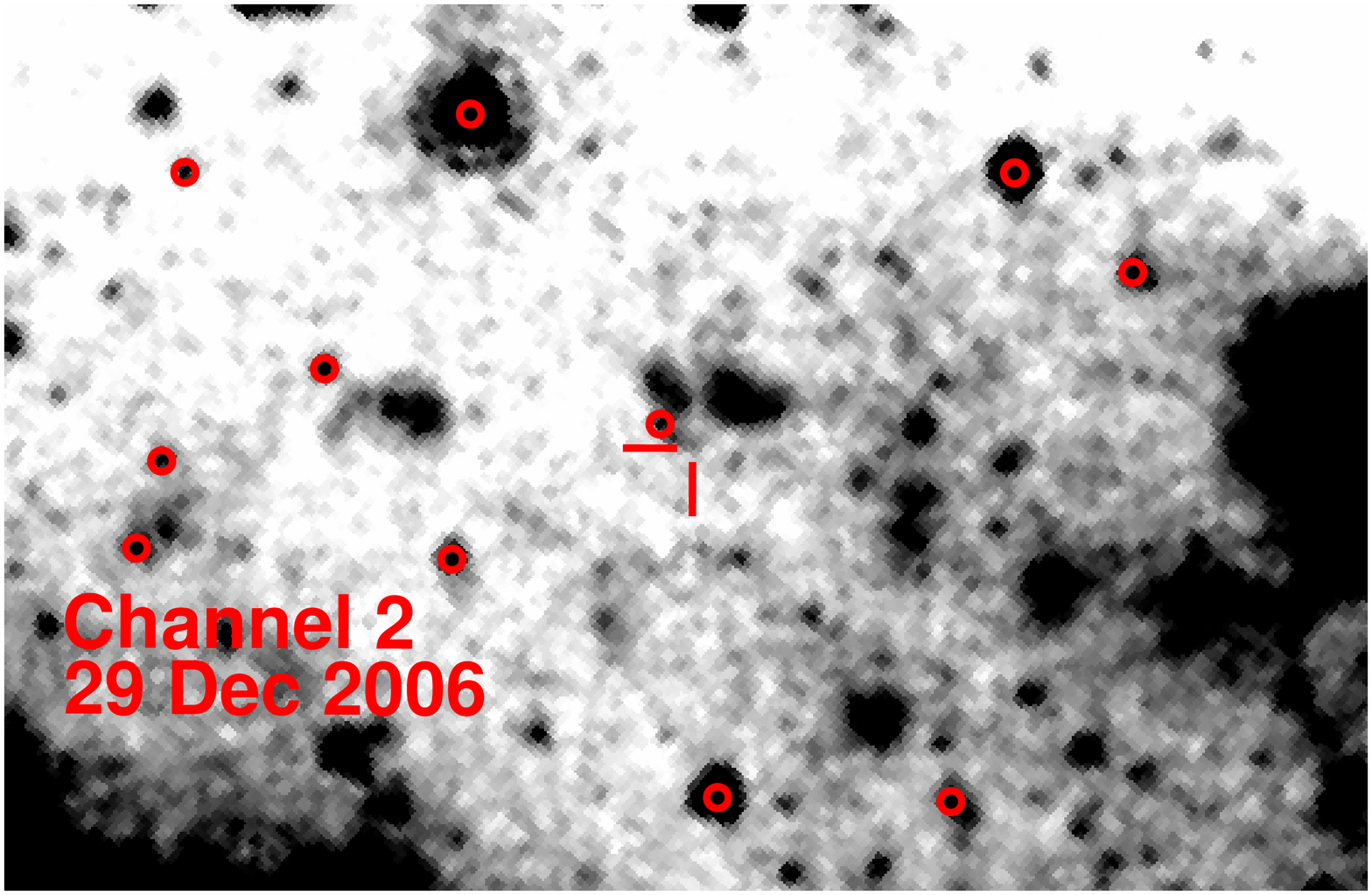}}
		\setlength{\fboxrule}{0.5pt}
		\caption{({\it Top}): \hst/WFC3 $F814W$ image from the top-left of \autoref{fig:hst} showing a 128\arcsec $\times$ 80\arcsec\ region with the location of SN~2017eaw denoted with red lines.  We circle $10$ sources used for relative astrometry with \spitzer\ imaging.  ({\it Middle}): \spitzer/IRAC Channel 1 imaging of NGC~6946 from 3 Jul. 2007.  We highlight the location of the pre-explosion mid-infrared counterpart of SN~2017eaw with red lines and circle the same $10$ common sources in the top panel.  ({\it Bottom}): Same as the middle panel, but for \spitzer/IRAC Channel 2 imaging from 29 Dec. 2006.}\label{fig:spitzer}
\end{figure}

We obtained \spitzer/IRAC exposures of NGC~6946 taken between 12 Sep. 2004 and 31 Mar. 2017 from the \spitzer\ Heritage Archive. The Basic Calibrated Data ({\tt bcd}) files were processed using {\tt MOPEX}, and each epoch was combined into a single frame with a scale of 0\arcsec.6 pixel$^{-1}$. SN~2017eaw was detected in a relatively crowded field and close to a cluster in the northern spiral arm of NGC~6946 (\autoref{fig:spitzer}). Using methods described in \citet{kilpatrick+18}, we performed unforced \textsc{IRAF}/{\tt daophot} photometry on all images using a point spread function (PSF) constructed empirically from isolated stars.  Each measurement was calibrated using zeropoints given in the IRAC instrument handbook for the cold or warm \spitzer\ mission, depending on the epoch of observation\footnote{\url{http://irsa.ipac.caltech.edu/data/SPITZER/docs/irac/iracinstrumenthandbook/17/}}.  We also calculate $3\sigma$ upper limits on the presence of a source at the location of SN~2017eaw by injecting fake stars with the empirical PSF and repeating our {\tt daophot} photometry.  We show example epochs for Channels 1 (3 Jul. 2007) and 2 (29 Dec. 2006) centered on the explosion site of SN~2017eaw in \autoref{fig:spitzer}.

\subsection{Spectroscopy}\label{sec:spectrum}

\begin{figure}
	\includegraphics[width=0.49\textwidth]{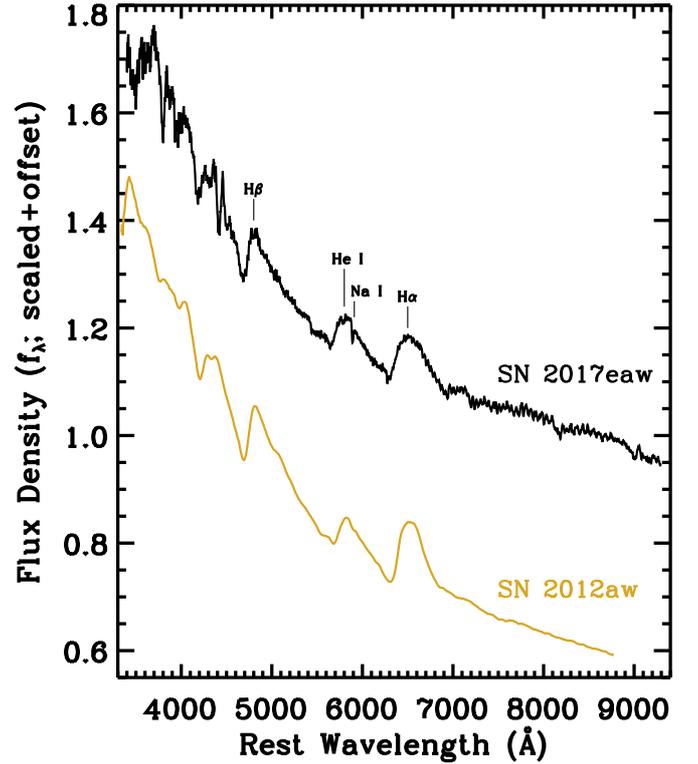}
	\caption{Our Faulkes North/FLOYDS spectrum of SN~2017eaw, which has been de-reddened for Milky Way extinction and the recessional velocity of NGC~6946 has been removed.  We note several spectroscopic features in this spectrum.  For comparison, we also plot a spectrum of SN~2012aw from \citet{dallora+14} and obtained 8~days after discovery of this SN.  We note the similar continuum shapes and spectroscopic features between both spectra.}\label{fig:spectra}
\end{figure}

We obtained a low-resolution spectrum of SN~2017eaw with the 2-m Faulkes Telescope North using the FLOYDS spectrograph on 19.52 May 2017.  This observation was facilitated by the Las Cumbres Observatory \citep[LCO;][]{brown+13} Global Telescope Network (NOAO--17AB, Program 12, PI Kilpatrick). We reduced and extracted the spectrum following standard procedures in {\tt IRAF}.  Using arc lamp spectra obtained immediately before the observation, we wavelength-calibrated the spectrum.  Finally, we performed flux calibration using a spectrum of the spectrophotometric standard LTT~4364 taken the previous night and in the same instrumental configuration.  Our final spectrum is shown in \autoref{fig:spectra}.

\section{RESULTS}\label{sec:results}

\subsection{Spectral Classification of SN~2017eaw}

Our spectrum of SN~2017eaw in \autoref{fig:spectra} has been de-reddened for Milky Way extinction and the recessional velocity of NGC~6946 \citep[$z=0.00133$;][]{epinat+08} has been removed.  The most noteable features are that this spectrum exhibits strong, blue continuum emission with relatively strong H$\alpha$ emission (full-width at half-maximum of 11,000~km~s$^{-1}$).  These spectral characteristics indicate that SN~2017eaw was a SN~II with a hot, optically thick photosphere similar to young SNe~II.  Given that we obtained our spectrum $5$~days after discovery, SN~2017eaw was likely very young at the time of observation.

We also compare SN~2017eaw to a spectrum of SN~2012aw at around 8~days after discovery \citep[as shown in \autoref{fig:spectra}; see also][]{dallora+14}.  The SN~2012aw spectrum has been de-reddened based on extinction estimates in \citep{dallora+14} and the recessional velocity of its host galaxy has been removed.  These two spectra are remarkably similar.  The fact that the continuum shapes are similar could suggest that we are fully accounting for Galactic, host, and circumstellar extinction in both objects, and thus that SN~2017eaw has very little host or circumstellar extinction, although this is still very uncertain.  There could also be a mismatch between the intrinsic continuum shapes.  We must consider other extinction indicators toward SN~2017eaw in order to separate the total extinction contribution from all sources.

NGC~6946 is a face-on spiral galaxy and there is no evidence for background emission or bright sources close to the position of SN~2017eaw that indicate it is embedded in a cluster or dense interstellar gas (\autoref{fig:hst}).  Typically, Na\I\ D absorption in spectra of the SN itself can provide a constraint on the host extinction to the progenitor source.  Although strong Na \I\ D absorption is present in our identification spectrum of SN~2017eaw, any host extinction would be blended with Milky Way extinction given our spectral resolution and the redshift to NGC~6946.  Therefore, this feature is likely dominated by the strong Milky Way extinction along this line of sight.  The equivalent width of the total Na \I\ D feature (which consists of blended Na\I\ D$_{1}$+D$_{2}$ in the identification spectrum) is $1.6\pm0.1$~\AA\, corresponding to Milky Way reddening of $E(B-V)=0.34\pm0.05$~mag assuming the relationship in \citet{poznanski+12}.  We assume this value for the total Milky Way and host reddening to SN~2017eaw and its progenitor system.  We also note that the Milky Way reddening to NGC~6946 is consistent with the lower limit of the value from Na \I\ D, implying that there could be effectively no host reddening.

\subsection{Astrometry Between SN~2017eaw and Pre-Explosion Imaging}\label{sec:astrom}

We performed relative astrometry between the post-explosion \hst/WFC3 image and the drizzled pre-explosion \hst/ACS and WFC3 images. For each pre-explosion frame, we identified 99--700 sources common to both the SN~2017eaw and pre-explosion image.  We then calculated and applied a WCS solution to the pre-explosion \hst\ image using the {\tt IRAF} tasks {\tt ccmap} and {\tt ccsetwcs}. We estimated the astrometric uncertainty of the new geometric projection in the pre-explosion \hst\ images by selecting random subsamples consisting of half of our common stars, re-calculating the geometric projection, and then determining the offsets between the remaining common stars. In this way, the astrometric uncertainty was generally $\sigma = 0.002$--$0.003$\arcsec (0.05--0.075 \hst/WFC3 pixels) in right ascension and declination.

We determined the position of SN~2017eaw in the post-explosion \hst/WFC3 $F814$ image to be $\alpha=20^{\rm h}34^{\rm m}44^{\rm s}.272$, $\delta=+60^{\circ}11^{\prime}36^{\prime\prime}.008$, which agrees with a single, unblended point source in all of the \hst\ images apart from ACS/$F658N$ where we do not detect any counterpart at $>3\sigma$.  There are no other detected point sources near the SN position and the background level is flat in this area indicating negligible contamination from faint coincident stars. The PSF parameters are also consistent with a point source.  The uncertainty on the position of this source is negligible, and so the total astrometric uncertainty is dominated by uncertainties in relative astrometry. In each \hst/WFC3 and ACS image, we do not detect any other point sources within a minimum of $0.176$\arcsec\ ($>$50$\sigma$) of the location of the location of the counterpart.  Therefore, we consider the detected sources to be a viable pre-explosion counterpart to SN~2017eaw.  We report the \hst\ photometry for this source in \autoref{tab:hst}.

\begin{table}
\begin{center}\begin{minipage}{3.3in}
      \caption{\hst\ Photometry of the SN~2017eaw Counterpart}\small
\begin{tabular}{@{}cccccc}\hline\hline
  Epoch\footnote{From discovery on 14.24 May 2017.}  & Instrument & Filter   & Exp. Time (s)   & Magnitude ($1\sigma$) \\ \hline
  $-$4671.21   & ACS/WFC    & $F814W$  & 120             & 22.550 (036) \\
  $-$4671.20   & ACS/WFC    & $F658N$  & 700             & $>$23.6      \\
  $-$459.48    & WFC3/IR    & $F128N$  & 2823.48         & 19.771 (032) \\
  $-$459.41    & WFC3/IR    & $F110W$  & 455.88          & 20.712 (012) \\
  $-$201.60    & WFC3/IR    & $F160W$  & 396.92          & 19.377 (007) \\
  $-$201.60    & WFC3/IR    & $F164N$  & 2396.93         & 19.109 (012) \\
  $-$199.50    & ACS/WFC    & $F606W$  & 2430            & 26.366 (049) \\
  $-$199.43    & ACS/WFC    & $F814W$  & 2570            & 22.825 (009) \\
\hline
\end{tabular}\label{tab:hst}
\end{minipage}
\end{center}
\end{table}

We estimate the probability of a chance coincidence in the \hst\ images by noting that there are roughly $4,000$--$8,000$ point sources with S/N$>3$ within a 20\arcsec\ radius of the location of SN~2017eaw in each image.  The 3$\sigma$ uncertainty ellipse for the \hst\ reference image has a solid angle of $\sim 2.5\times10^{-4}~\text{arcsec}^{2}$, which implies that 0.16\% of the region within 20\arcsec\ of SN 2017eaw is similarly close (within 3$\sigma$) to any point source. This value is roughly the probability of a source aligning with the position of SN~2017eaw by chance.

The source appears to have decreased in $F814W$ luminosity by 30\% from $12.7$ to $0.6$~yr before core collapse.  This is a significant change of $7.8\sigma$, implying significant variability in the SED of this source before SN~2017eaw underwent core-collapse.

We also performed relative astrometry between the \hst/WFC3 image of SN~2017eaw and \spitzer/IRAC photometry in order to determine whether there was any mid-infrared source consistent with being the progenitor system of SN~2017eaw.  Because the point source full-width at half-maximum (FWHM) is much larger and the signal-to-noise per source is much lower in the \spitzer/IRAC images relative to the \hst\ images, there were significantly fewer sources to anchor \spitzer/IRAC to the \hst\ image of SN~2017eaw. In each image, we typically used $7$--$15$ point sources to calculate an astrometric solution.  The astrometric uncertainties in the \spitzer/IRAC solutions were typically $\sigma=0.24$~pixels, or $0.144\arcsec$ in both directions.  In \autoref{fig:spitzer}, we show example \spitzer/IRAC images in Channels 1 and 2 along with the \hst/WFC3 image of SN~2017eaw.  The $10$ sources used for relative astrometry are circled in red in each image, which correspond to the {\tt daophot} positions in the \spitzer/IRAC images and {\tt dolphot} positions in the \hst\ image.

In each \spitzer/IRAC image, there is at most one point source consistent with being the progenitor star of SN~2017eaw.  No other \hst\ sources are coincident with the \spitzer\ source.  The position of this source agrees with the \hst\ position to within the 1$\sigma$ uncertainties and the closest point source identified in any image is $3.4$\arcsec\ ($23.6\sigma$) away from this source.  Following the method outlined above for \hst\ imaging, we estimate the chance coincidence in \spitzer/IRAC imaging within a 20\arcsec\ radius of this source to be $2.3\%$ or smaller per image.  Thus, it is unlikely that this source is a chance coincidence and is therefore likely to be the mid-infrared pre-explosion counterpart to SN~2017eaw.  We detect this source in every Channel 1 and 2 image for which we have data but not in any of the Channel 3 or 4 images, where we place upper limits on the flux density of any such source.  We report all detections and upper limits in \autoref{tab:spitzer}.

\begin{table}
\begin{center}\begin{minipage}{3.3in}
      \caption{\spitzer\ Photometry of the SN~2017eaw Counterpart}\small
\begin{tabular}{@{}cccccc}\hline\hline
 Epoch\footnote{From discovery on 14.24 May 2017.}  & Channel 1    & Channel 2    & Channel 3 & Channel 4  \\
               & ($\mu$Jy)    & ($\mu$Jy)    & ($\mu$Jy) & ($\mu$Jy)  \\ \hline
$-$4719.98     & 17.7$\pm$3.8 & 11.0$\pm$1.3 & $<$23.5   & $<$67.8    \\
$-$4626.45     & 18.6$\pm$3.5 & 11.1$\pm$1.3 & $<$19.1   & $<$62.1    \\
$-$4552.00     & 17.7$\pm$3.8 & 11.4$\pm$1.1 & $<$30.0   & $<$77.6    \\
$-$4315.51     & 17.1$\pm$3.6 & 10.1$\pm$1.5 & $<$27.9   & $<$64.8    \\
$-$4151.85     & ---          & ---          & ---       & $<$86.8    \\
$-$3927.24     & ---          & ---          & $<$29.8   & ---        \\
$-$3820.83     & ---          & 12.5$\pm$1.6 & ---       & $<$75.5    \\
$-$3788.71     & ---          & 11.9$\pm$2.7 & ---       & $<$82.7    \\
$-$3601.77     & 17.9$\pm$3.8 & ---          & ---       & ---        \\
$-$3425.67     & ---          & 10.6$\pm$1.5 & ---       & $<$80.8    \\
$-$3393.89     & ---          & 8.6$\pm$1.7  & ---       & $<$107.4   \\
$-$3220.95     & 14.9$\pm$3.4 & ---          & $<$86.8   & ---        \\
$-$2837.75 &      14.7$\pm$       2.0 &--- &---&--- \\
$-$2685.13 &--- &      10.1$\pm$       0.8 &---&--- \\
$-$2465.17 &      12.6$\pm$       2.0 &--- &---&--- \\
$-$2116.88 &      16.4$\pm$       3.9 &--- &---&--- \\
$-$2111.92 &      17.6$\pm$       2.1 &--- &---&--- \\
$-$1928.68 &--- &       9.9$\pm$       0.9 &---&--- \\
$-$1365.60 &      19.8$\pm$       2.1 &--- &---&--- \\
$-$1226.16 &--- &       9.3$\pm$       1.6 &---&--- \\
$-$1181.13 &--- &       9.4$\pm$       0.8 &---&--- \\
$-$1144.19 &      15.2$\pm$       2.0 &      11.2$\pm$       0.9 &---&--- \\
$-$970.55 &      16.7$\pm$       1.9 &      11.9$\pm$       0.9 &---&--- \\
$-$941.18 &      16.7$\pm$       2.0 &      11.0$\pm$       0.8 &---&--- \\
$-$832.76 &--- &      10.9$\pm$       1.6 &---&--- \\
$-$619.49 &      15.8$\pm$       2.0 &      13.4$\pm$       0.9 &---&--- \\
$-$613.84 &      16.7$\pm$       1.9 &      12.3$\pm$       0.8 &---&--- \\
$-$604.85 &      16.4$\pm$       2.1 &      12.4$\pm$       0.8 &---&--- \\
$-$592.75 &      15.5$\pm$       1.9 &      13.4$\pm$       0.7 &---&--- \\
$-$535.30 &      17.5$\pm$       1.8 &      14.4$\pm$       0.7 &---&--- \\
$-$506.86 &      16.5$\pm$       1.9 &      13.7$\pm$       0.8 &---&--- \\
$-$213.44 &      16.6$\pm$       1.9 &      13.9$\pm$       0.8 &---&--- \\
$-$134.31 &      17.7$\pm$       1.9 &      12.9$\pm$       0.8 &---&--- \\
$-$42.76 &      17.9$\pm$       2.0 &      13.4$\pm$       0.9 &---&--- \\ \hline
\end{tabular}\label{tab:spitzer}
\end{minipage}
\end{center}
\end{table}

\subsection{Mid-Infrared Light Curves of the Progenitor System}\label{sec:mirlc}

\begin{figure}
	\includegraphics[width=0.475\textwidth]{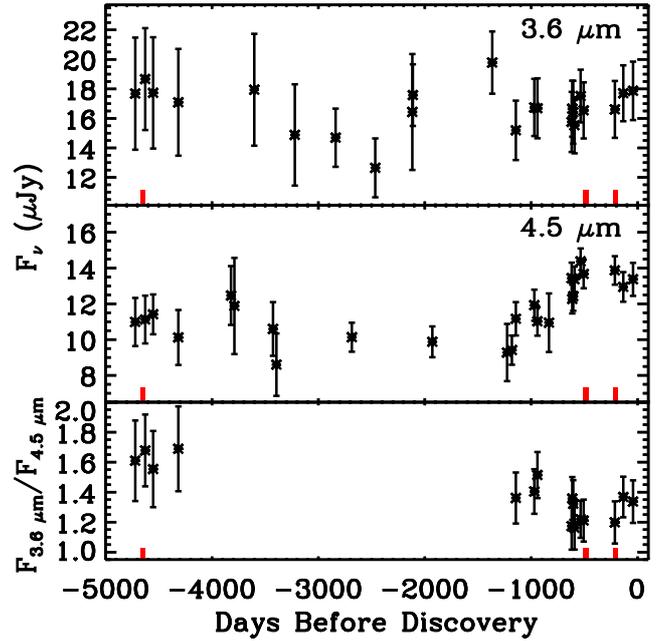}
	\caption{Full 3.6~$\mu$m ({\it Upper}) and 4.5~$\mu$m ({\it Middle}) light curves of the SN~2017eaw counterpart.  The source does not appear to be significantly variable at 3.6~$\mu$m, but it exhibits a notable rise in $4.5~\mu$m emission starting around 1000 days before the discovery.  We also plot the ratio of the emission in the two \spitzer\ bands where they are contemporaneous.  The source is significantly brighter in 4.5~$\mu$m emission relative to 3.6~$\mu$m (i.e., redder/cooler) during the last several years before core-collapse relative to $\sim$12~yr before core-collapse.  We note the three epochs of contemporaneous \hst\ imaging using red bars in each panel.}\label{fig:lc}
\end{figure}

In \autoref{fig:lc}, we plot the {\it Spitzer}/IRAC Channels 1 and 2 (3.6 and 4.5$\mu$m) light curves of the SN~2017eaw counterpart.  The light curve indicates that the mid-infrared source was persistent for over $\sim$13~yr prior to core collapse.  The median and standard deviation of the total {\it Spitzer} light curves are 16.7 and 1.5~$\mu$Jy at 3.6$\mu$m and 11.4 and 1.6$~\mu$Jy at 4.5$\mu$m.  At 3.6$\mu$m, 96\% of the of the data are within 1$\sigma$ of the median and 100\% are within 2$\sigma$, which indicates that the data are consistent with measurements of a single value with standard errors and thus exhibited no variability.

At the same time, only 64\% of the 4.5$\mu$m data are within 1$\sigma$ of the median while 76\% are within 2$\sigma$.  Statistically, this finding indicates that there is likely some variability in this waveband.  Combined with the apparent lack of variability at 3.6$\mu$m, the source appears to be changing both in overall mid-infrared color and luminosity.  This change is qualitatively apparent in the light curve from \autoref{fig:lc}, where it appears to increase in 4.5$\mu$m luminosity $1200$--$500$~days before discovery of SN~2017eaw.  At the same time, the ratio of 3.6$\mu$m to 4.5$\mu$m emission decreases (i.e., the source becomes redder/cooler).

There may still be variability over relatively short timescales.  Within the uncertainties of each mid-infrared epoch, the only significant change we detect is the general increase in $4.5\mu$m flux from $1200$--$200$~days before discovery.  The data in this span of time probe timescales as short as $6$~days, which implies that within our photometric precision, there were no significant changes on these timescales.  However, our uncertainties are typically $\sim$10\% the value of each measurement, and so we are not sensitive to variability comparable to or smaller than this scale.

This trend suggests that in the mid-infrared, the source was variable within $\sim$4~yr of core-collapse.  If the underlying source is a RSG with a dusty wind \citep[similar to Galactic analogs;][]{massey+05}, then it is likely that the 4.5$\mu$m behavior was driven by dust production.  Perhaps this system had an enhanced mass-loss episode leading to a higher density of dust in its circumstellar environment.  The mid-infrared emission would have become more optically thick and cooler on a relatively short timescale as the infrared SED shifted to longer wavelengths.  This hypothesis is in agreement with the decrease in optical luminosity inferred from $F814W$.

However, the exact nature of the underlying source powering this dust production is less clear from the \spitzer\ light curve.  It is possible that the associated optical emission from this source would also have been variable as the RSG underwent an enhanced mass-loss episode.  Isolating the SED at a specific epoch in time is essential to determine the intrinsic luminosity of the SN~2017eaw progenitor star, and thus determine its initial mass.

\subsection{Evolution of the Spectral Energy Distribution of the Pre-Explosion Source}\label{sec:dust}

\begin{figure}
	\includegraphics[width=0.475\textwidth]{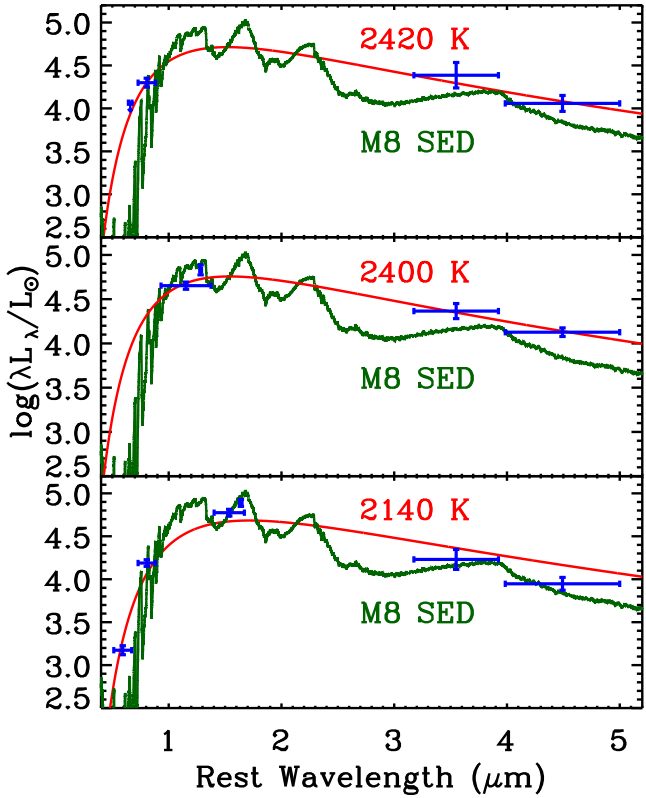}
	\caption{The optical-infrared spectral energy distribution of the SN~2017eaw counterpart over three epochs (blue points; earliest to latest from top to bottom) for which we have \hst\ coverage.  The data are corrected for Milky Way reddening, shifted to the rest-frame of NGC~6946, and shifted to the assumed distance to SN~2017eaw.  We fit the SED in each epoch with a pure blackbody profile (red) with no additional source of extinction or emissivity and note the temperature of the profile in each panel.  Under the assumption that this source is an extremely cool supergiant with little or no excess mid-infrared emission, we also fit the SED to a M8 star (green) based on a MARCS photospheric model \citep{gustafsson+08}.}\label{fig:sed}
\end{figure}

Under the assumption that the pre-explosion source is mostly dominated by thermal emission from a dust shell or stellar photosphere, we track the evolution of that source over three epochs for which we have \hst\ and \spitzer\ data to analyze the optical-infrared variability of that source.  These epochs correspond to 29 Jul 2004 to 11 Sep 2004, 23 Dec 2015 to 9 Sep 2016, and 12 Oct 2016 to 26 Oct 2016 (roughly 12.7, 1.3, and 0.6~yr before discovery or core-collapse, respectively).  Our fits to the overall SED in these three epochs is shown in \autoref{fig:sed}.  For simplicity, we fit blackbody emission (i.e., with emission efficiency $Q=1$ at all wavelengths).

The best-fitting SED is a thermal source with $\log(L/L_{\odot})=4.61\pm0.21$, $4.70\pm0.24$, and $4.80\pm0.24$ and $T=2540\pm160$~K, $2360\pm200$~K, and $2070\pm220$~K in the three epochs, respectively.  These values correspond to a photospheric radius of $1000$, $1300$, and $2000~R_{\odot}$.  Overall, these fits suggest the underlying source is cooling and its photosphere is expanding with time, although the luminosity is consistent with being constant.

The simple blackbody model is only a relatively good fit to the full optical to mid-infrared SED in the first epoch of data, which is also where the predicted temperature is hottest and photospheric radius is smallest.  Indeed, $1000~R_{\odot}$ is comparable to the radius of many $15$--$16~M_{\odot}$ RSGs \citep[based on fits to Mesa Isochrone \& Stellar Tracks (MIST) models;][]{paxton+11,paxton+13,paxton+15,dotter+16,choi+16}\footnote{\url{http://waps.cfa.harvard.edu/MIST/}}.

Suppose that we are seeing the photosphere of a single, ultra-cool supergiant.  Detailed analysis of the SEDs and TiO bands observed from Galactic RSGs indicates that the coolest intrinsic temperatures are $\approx3450$~K \citep[for M5~I stars; see, e.g.,][]{levesque+05}.  Even if the pre-explosion source were an anomalously cool supergiant with a photosphere approaching an unprecedented value of $2600$~K, model photospheres \citep[e.g., from MARCS][]{gustafsson+08} with those temperatures fail to simultaneously reproduce the observed optical and mid-infrared photometry (see M8~I fits in \autoref{fig:sed}).  The SED is either too bright in the mid-infrared, suggesting some excess source of cool emission (e.g., dust), or too bright in the optical suggesting that we are in fact seeing a hotter photosphere (e.g., from a warmer star).  It is more likely that the intrinsic SED has two components consisting of an optical stellar component reddened by circumstellar material and a mid-infrared component from reprocessed emission.

Under the assumption that the SED is dominated by reddened optical emission and mid-infrared dust emission, this would seem to point to a relatively cool star, such as a RSG with a slow, dusty wind that is reddening the intrinsic SED.  Overall, the SED is similar to Galactic RSGs, whose SEDs peak around $1.3$--$1.6\mu$m and exhibit spectroscopic signatures of infrared dust emission \citep{verhoelst+09}.  However, we must consider the possibility that the intrinsic SED of the pre-explosion counterpart is a significantly hotter source (e.g., a yellow or blue supergiant) that is even more heavily reddened.  Below, we perform a physically-motivated analysis of the SED assuming it represents a star with an arbitrary temperature with circumstellar extinction and dust emission.

\subsection{Detailed Stellar SED and Dust Models}\label{sec:models}

The optical through mid-infrared SED of the SN~2017eaw progenitor star spanning $214$--$200$~days before discovery of the SN offers an unprecedented opportunity to explore the exact spectral type and characteristics of a SN~II progenitor star within the final months before core-collapse.  Therefore, we restrict our analysis below to only the \hst/WFC3 data from Julian Date 2457685.64--2457687.81 and the \spitzer/IRAC data from 2457673.81 (i.e., plotted in the bottom panel of \autoref{fig:sed}).  These data provide a ``snapshot'' of the optical to mid-infrared SED of the SN progenitor star over a very narrow window of time and minimise any systematic uncertainties associated with pre-SN variability on timescales longer than $14$~days.  

As we discuss in \autoref{sec:mirlc}, we are not sensitive to mid-infrared variability on scales $\sim$10\% or smaller.  Moreover, we have no constraint on optical or near-infrared variability apart from the 30\% decrease in $F814W$ flux over roughly $12$~yr.  We cannot rule out the possibility that the SN~2017eaw progenitor system is somewhat variable over these $14$~days, and so we increase our overall uncertainties by a factor of $2$ for the following analysis.

In order to fit these data and investigate the exact SED of the progenitor star, we followed a procedure similar to \citet{kochanek+12}, who analyzed the SED of the progenitor star of SN~2012aw.  We began with SEDs derived from MARCS $15~M_{\odot}$ RSG models of standard composition and spherical geometry \citep[for full descriptions of MARCS models, see][]{gustafsson+08}.  We considered only models at the metallicity of NGC~6946 and with a turbulence paramter of $5$~km~s$^{-1}$.  Otherwise, we investigated every available MARCS model at a fixed mass, chemical class, geometry, metallicity, and turbulence parameter, but with the full range of available surface gravities ($\log~g=-0.5$ to $1.0$ in steps of $0.5$) and surface temperatures ($T=2500$--$4000$~K in steps of 100~K as well as $4250$~K and $4500$~K).  In order to expand the range of temperatures in our analysis, we supplemented this set of models with models of hotter ($5000$--$8000$~K in steps of $1000$~K) and cooler ($2600$--$3200$~K in steps of $200$~K) photospheres with the same chemical class, geometry, metallicity, and turbulence parameter, but based on a $5~M_{\odot}$ model and with a range of surface gravities.

To reduce the computational complexity of our analysis, we smoothed each SED of each MARCS model by a factor of $13$ from the default resolution of $\lambda/\Delta\lambda=20,000$ to $\approx$1540.  As in \citet{kochanek+12}, we linearly interpolated between models with different temperatures in order to fit intermediate temperature values.  Otherwise, we restricted our analysis to only photospheres with temperatures between $2600$ and $8000$~K.

To account for optical and infrared extinction due to a shell of dust, we applied a circumstellar extinction law to each MARCS SED, which was calculated from DUSTY models by \citet{kochanek+12}.  We repeated our fitting process for four different types of circumstellar extinction (graphitic/silicate dust grains and $R_{out}/R_{in}=2$ and $10$, where $R_{out}$ and $R_{in}$ are the outer and inner radii of the dust shell) and the four different values of $\log g$ without parameterizing over these quantities (i.e., we tested the fit 16 times for all combinations of these dust types and surface gravities).  In general, we found the best fits using graphitic models with $R_{out}/R_{in}=2$ and stellar SEDs with $\log g = 1.0$.

We tested our fits by comparing the specific luminosity ($L_{\lambda}$) of SN~2017eaw counterpart in each filter, that is the observed flux density ($f_{\lambda}$) corrected for the total Milky Way and host extinction extinction from above ($A_{MW+H,\lambda}$) and at the distance of NGC~6946 ($d$) such that $L_{\lambda} = 4\pi d^{2} f_{\lambda} 10^{0.4 \times A_{MW+H,\lambda}}$.  We compared these values to the modeled SED $L_{o,\lambda}$ convolved through each \hst\ and \spitzer\ filter transmission function.  This model SED is calculated as 

\begin{equation}
L_{o,\lambda}=\frac{\int_{0}^{\infty} \lambda R_{\lambda} (L_{\star,\lambda} 10^{-0.4 \times A_{\lambda}(\tau_{V})}+L_{d,\lambda}) d\lambda}{\int_{0}^{\infty} \lambda R_{\lambda} d\lambda}.
\end{equation}

\noindent Here, $L_{\star,\lambda}$ is the scaled, interpolated MARCS model for a certain temperature and surface gravity, $A_{\lambda}(\tau_{V})$ is the total extinction due to dust calculated from the optical depth of the dust shell in $V$-band ($\tau_{V}$) and for each wavelength of the MARCS SED $\lambda$ using Table~3 in \citet{kochanek+12}, $L_{d,\lambda}$ is the total luminosity due to dust emission, and $R_{\lambda}$ is the filter transmission function.  

The dust emission is reprocessed optical light from the obscured star, which illuminates and heats the dust.  Therefore, the total dust luminosity is dependent on the luminosity of the underlying star as well as the fraction of that light that is absorbed by the dust (i.e., it is related to $A_{\lambda}$).  Here, we parameterize the fraction of the total luminosity that is absorbed by dust as 

\begin{equation}
f=\frac{\int_{0}^{\infty} (L_{\star,\lambda} - L_{\star,\lambda} 10^{-0.4 A_{\lambda}}) d\lambda}{\int_{0}^{\infty} L_{\star,\lambda} d\lambda}.
\end{equation}

\noindent Thus, the total dust luminosity is related to the intrinsic luminosity of the star as $L_{d} = f L_{\star}$.

Following analysis in \citet{kilpatrick+18} \citep[see also][]{fox+10,fox+11}, we assume the infrared dust emission is optically thin.  Therefore, $L_{d,\lambda} = L_{0} B_{\lambda}(T_{d}) \kappa_{\lambda}$ where $B_{\lambda}(T_{d})$ is the wavelength-dependent Planck function for a dust temperature $T_{d}$ and $L_{0}$ is a normalization constant such that $\int_{0}^{\infty} L_{d,\lambda} d\lambda = L_{d}$ as described above.  We take $\kappa_{\lambda}$ from fig. 4 of \citet{fox+10} for dust grains with diameter $0.01~\mu$m, which is roughly consistent with the weighted-average ($\approx 0.0083~\mu$m) of the \citet{mathis+77} power-law grain size distribution used in \citet{kochanek+12}.  Furthermore, as \citet{fox+10} note, below $1~\mu$m the dust grain size has very little effect on the infrared opacities.

Thus, the full model is parameterized over the total luminosity of the star ($L_{\star}=\int_{0}^{\infty} L_{\star,\lambda} d\lambda$), the temperature of the interpolated MARCS model ($T_{\star}$), the optical depth of the dust in $V$-band ($\tau_{V}$), and the temperature of the dust ($T_{d}$).  We fit these four parameters to the six \hst/ACS and \spitzer/IRAC specific luminosities from the bottom panel of \autoref{fig:sed} (i.e., with two degrees of freedom) using a Monte Carlo Markov Chain (MCMC) method by minimising $\chi^{2} = (L_{\lambda} - L_{o,\lambda})^{2} / (\sigma_{L_{\lambda}})^{2}$ summed over each data point.  Here, $\sigma_{L_{\lambda}}$ is the uncertainty for specific luminosity in each filter (including the uncertainties quoted above for photometry and Milky Way extinction).  Although we account for distance uncertainty in our final stellar luminosity, the other parameters have no dependence on distance (we simply scaled the input flux densities by our preferred distance), and so we did not include distance uncertainty in our MCMC.

\begin{figure*}
	\includegraphics[width=0.48\textwidth]{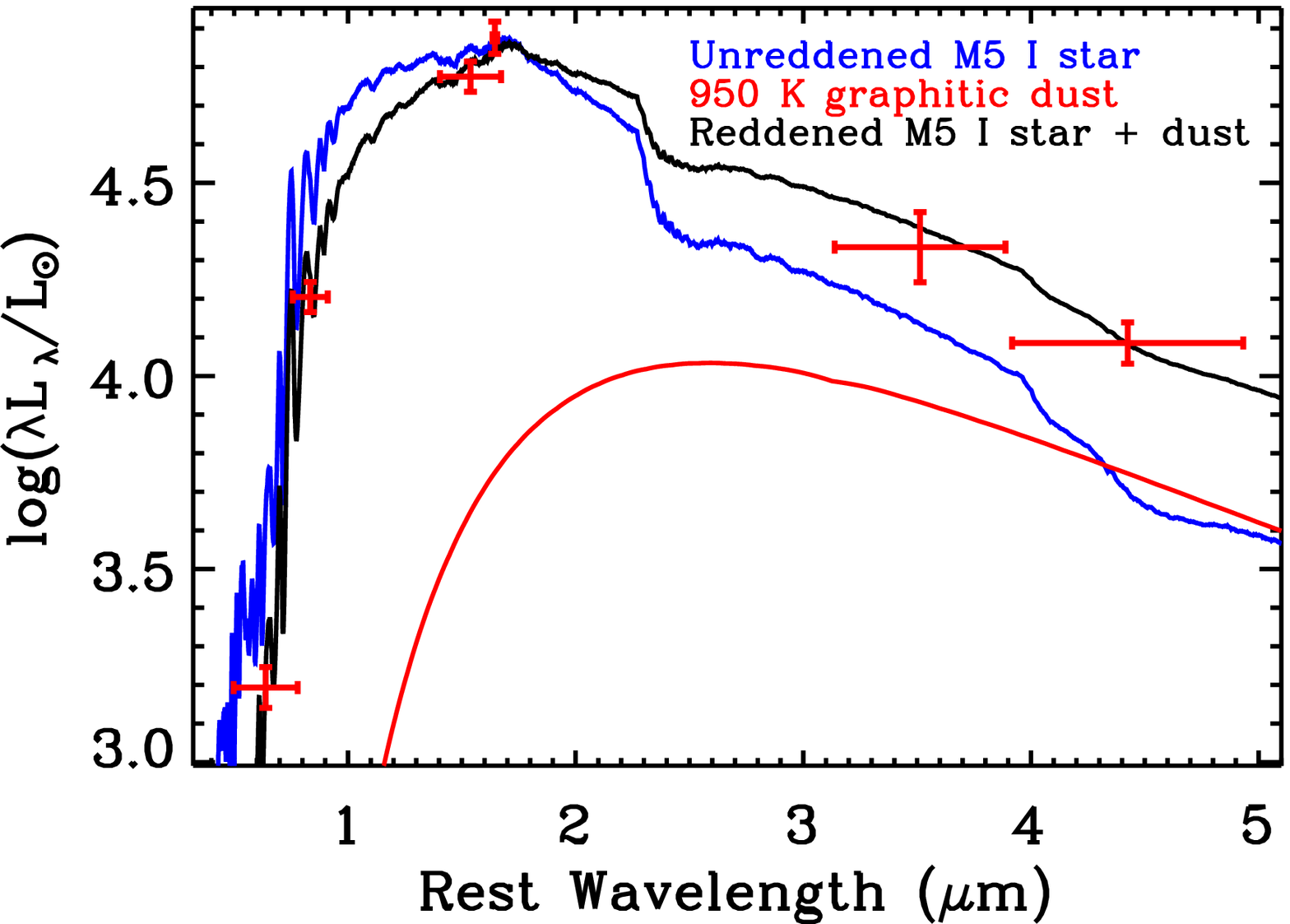}
	\includegraphics[width=0.48\textwidth]{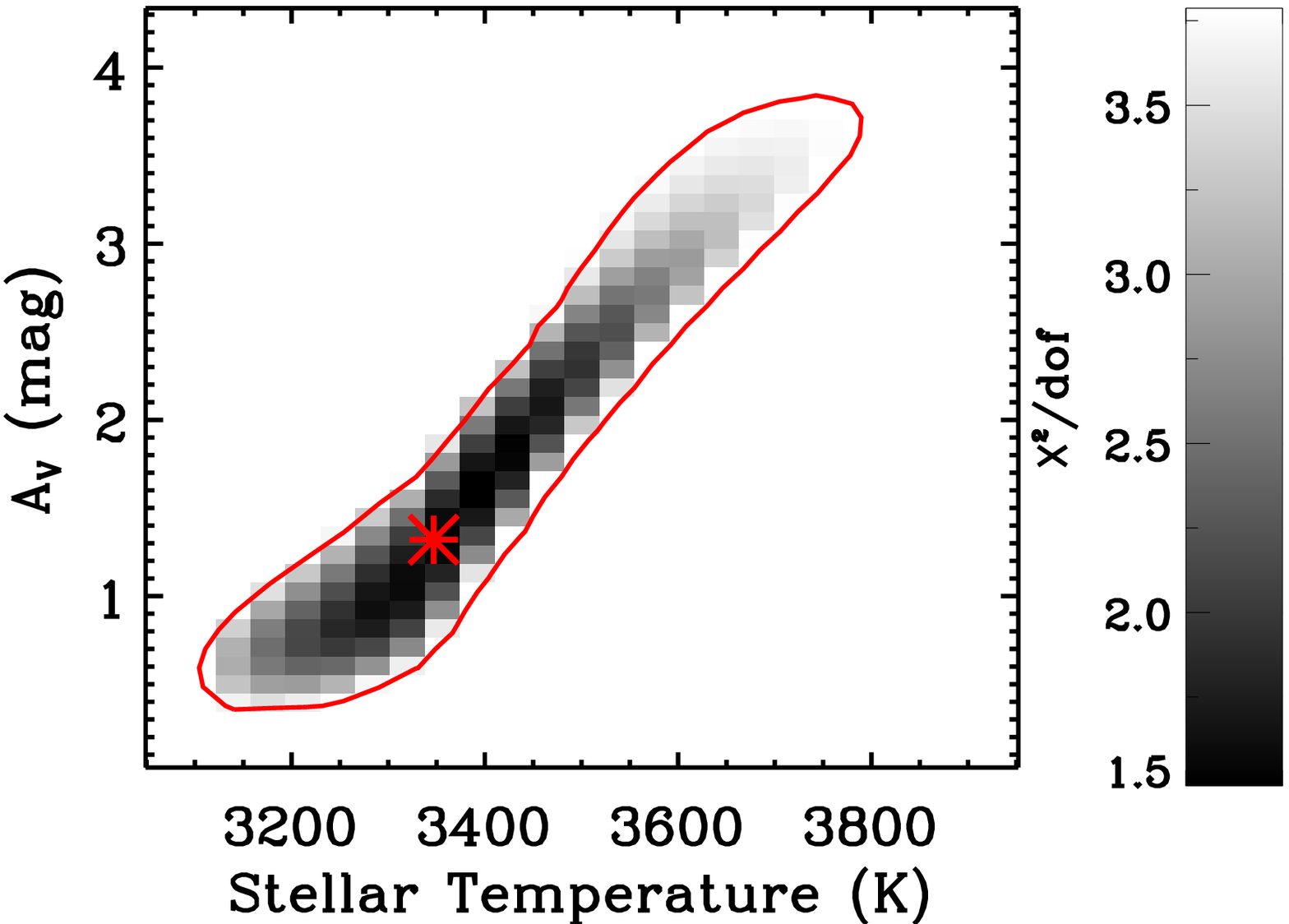}
	\caption{({\it Left}) The optical-infrared SED of the SN~2017eaw counterpart from Julian Date 2457685.64--2457687.81.  The data are corrected for Milky Way reddening, shifted to the rest-frame of NGC~6946, and shifted to the assumed distance to SN~2017eaw.  We fit the SED in each epoch with a MARCS RSG model (blue curve) that is reddened according to the circumstellar extinction law from \citet{kochanek+12}.  We combine this SED with $0.01~\mu$m graphitic dust (red curve). The total SED, which is convolved with the relevant \hst\ and \spitzer\ filter functions and fit to the observed data, is shown as a black curve.  ({\it Right}) $\chi^{2}/\text{dof}$ parameter estimate of the intrinsic stellar temperature and extinction due to circumstellar material \citep[in $A_{V} = 0.79 \tau_{V}$ for graphite as in][]{kochanek+12}.  The red star marks the best-fitting parameters while the red line marks the full $1\sigma$ uncertainty range (i.e., $\chi^{2}/~\text{dof} = \chi_{min}^{2}/~\text{dof} + 2.3$).  There is a large degeneracy between these two values, although we can reasonably constrain the stellar temperature between $3100$ and $3800$~K.}\label{fig:fit}
\end{figure*}

Our best-fitting model to the specific luminosities is shown in \autoref{fig:fit}.  The MCMC converged with a final $\chi^{2}/\text{dof} = 1.5$.  We estimated our uncertainties by varying each parameter while fixing the other three to their best-fitting values and determining where $\chi^{2}/\text{dof}$ increased by $2.3$ (i.e., the 68\% confidence interval for a $\chi^{2}$ distribution with 2 degrees of freedom).  The luminosity is robustly predicted to be $\log(L/L_{\odot})=4.9\pm0.2$ regardless of the intrinsic stellar type or dust properties, which is consistent with the simple blackbody fits above.  

There is significant degeneracy between $\tau_{V}$ and the stellar temperature \citep[see \autoref{fig:fit}; also discussed in][]{kochanek+12,fraser+12}.  Models with hotter temperatures can produce good fits to the observed data if the circumstellar extinction is higher.  This degeneracy can lead to tight or loose constraints on the temperature and circumstellar extinction, although these constraints have systematic uncertainties that are specific to the circumstellar extinction model.  We show the full range of this degeneracy (within the formal $1\sigma$ uncertainties) for our best-fitting circumstellar extinction \citep[in $A_{V}\approx0.79 \tau_{V}$ as in][]{kochanek+12} and stellar temperature.  To a lesser extent, the dust temperature is also degenerate with these quantities; in cases where the stellar temperature is low, the dust shell is relatively low luminosity and cool, and it is luminous and hot for a hot star.

These constraints imply a relatively cool best-fitting stellar temperature of $T_{\star}=3350\substack{+450\\-250}$~K, $V$-band optical depth of $\tau_{V}=1.6\substack{+3.3\\-1.2}$, and a dust temperature of $T_{d}=950\substack{+450\\-400}$~K.  The implied dust luminosity varies from $\log(L_{d}/L_{\odot})=4.1\substack{+0.3\\-0.4}$ with a radius $R_{in}=4000\substack{+4000\\-1300}~R_{\odot}$.  The radius of this dust shell is approximately $5$~times the photospheric radius of the best-fitting stellar model.

\subsection{Physical Properties of the RSG and Circumstellar Material}

\begin{figure}
	\includegraphics[width=0.475\textwidth]{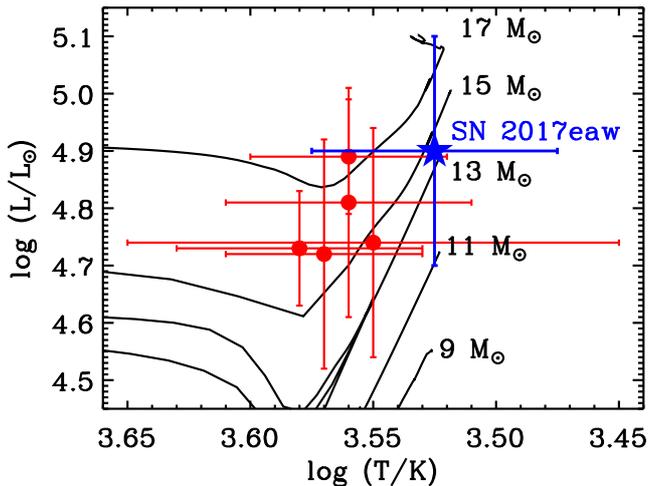}
	\caption{A Hertzsprung-Russell diagram showing the location of our best-fitting model to the SN~2017eaw counterpart (blue star) roughly $200$~days before core-collapse.  For comparison, we also show the locations of several SNe~II progenitor stars from \citet{smartt15} (red circles).  We also plot MIST evolutionary tracks for $9$--$17~M_{\odot}$ stars.}\label{fig:tracks}
\end{figure}

We compared the total luminosity and temperature of the model star to evolutionary tracks from MIST.  Examining only MIST tracks at Solar metallicity and with rotation velocity $v_{rot}/v_{crit}=0.4$, we find the best fits to evolutionary tracks with initial masses $13\substack{+4\\-2}~M_{\odot}$ (\autoref{fig:tracks}).

At the same time, $\tau_{V}$ can be used to constrain the mass of circumstellar material.  Using equation (2) in \citet{kochanek+12}, we assume a total mass in the dusty wind of $M = 4 \pi \tau_{V} \kappa_{V}^{-1} R_{out} R_{in}$ and a mass-loss rate $\dot{M} = 4 \pi \tau_{V} \kappa_{V}^{-1} v_{w} R_{in}$.  We assume a wind velocity $v_{w} \approx 10$~km~s$^{-1}$ and we derive our visual opacity $\kappa_{V}$ from the above dust model.

For the circumstellar material around SN~2017eaw, we find that the total dust mass at the epoch $200$~days before core-collapse is $(2\pm0.5)\times10^{-5}~M_{\odot}$ and the mass-loss rate is $9\times10^{-7}~M_{\odot}~\text{yr}^{-1}$.  The former quantity assumes $R_{out}/R_{in}=2$, although the outer radius of the dust shell is poorly constrianed by our model (we tested only two possible geometries in \autoref{sec:models}), and it could reasonably be much larger than $2\times R_{in}$.  Thus, we consider this dust mass to be a lower limit on the total mass of material.

The mass-loss rate we derive precisely follows the de Jager rate for a star at the derived luminosity of $\log L = 4.9$ \citep{woodhams+93,huggins+94,reimers+08,mauron+10,braun+12}.  However, the wind velocity is completely unconstrained by our data and could be a few times larger \citep[e.g., $30$--$50$~km~s$^{-1}$, which is the wind velocity for high-luminosity RSGs such as VY~CMa or NML~Cyg;][]{knapp+82,decin+06}.  Detailed observations of SN~2017eaw will be necessary to better constrain this parameter.

However, detailed hydrodynamic and pan-chromatic studies of mass loss in RSGs demonstrate that many stars lie below the de Jager prescription depending on the mass-loss tracer and the assumed level of clumping and metallicity \citep[][]{mauron+10}.  \citet{smith+14} claim that mass-loss rates are overestimated because they are derived from H$\alpha$, infrared, or radio luminosities assuming a homogeneous wind, and clumping can produce similar luminosities with less material.  Our measurement of the SN~2017eaw progenitor star mass-loss rate makes a similar assumption (i.e., we assume a uniform wind density $\rho \sim r^{-2}$), and we recover a mass-loss rate that is similar to or larger than Galactic analogs.  The level of clumping in the wind can imply a significantly lower mass-loss rate.

\section{DISCUSSION}\label{sec:discussion}

We find that the SN~2017eaw progenitor star had an initial mass close to $13\substack{+4\\-2}~M_{\odot}$, which agrees with the mass derived in \citet{atel10378}. This quantity is mostly dependent on the luminosity of the SN~2017eaw progenitor star and thus the distance and intrinsic SED of this source.  If we use the closer distance to NGC~6946 \citep[5.7~Mpc from, e.g.,][]{sahu+06}, then the implied luminosity would be $\log(L/L_{\odot})=4.7\pm0.2$ and the initial mass would be $11\substack{+3\\-2}~M_{\odot}$. Overall, the SN~2017eaw progenitor star is consistent with the observed distribution of SN~II progenitor stars, where it is typically found that they are RSGs in the $3400$--$4000$~K range with luminosities below $\log(L/L_{\odot})=5.2$.  The photometric evolution of SN~2017eaw \citep{tsvetkov+18}, which involved a long plateau phase, and broad lines of H$\alpha$ in early-time spectra of SN~2017eaw \citep[\autoref{fig:spectra} and][]{atel10374} all support the conclusion that the progenitor star was a RSG with an extended hydrogen envelope.

Our preferred mass is high for a SN~II progenitor system \citep[in the upper 76th percentile of the distribution in][]{smartt15}.  We considered the probability that all of the current sample of SN~II progenitor stars with reported initial mass estimates would all have $8~M_{\odot} < M_{init} < 17~M_{\odot}$.  Assuming a Salpeter initial mass function, this is a 0.26\% probability for $14$ sources.  Although this is a relatively crude estimate of the likelihood of an upper limit on mass compared to \citet{smartt15} and \citet{davies+18}, this analysis suggests that the current sample is consistent with an upper mass limit assuming the masses are all accurate.

The evidence for circumstellar extinction around the SN~2017eaw progenitor star supports the conclusion that some progenitor mass estimates are low.  In particular, the luminosities of candidate progenitor stars with only a single band of pre-explosion imaging have large systematic uncertainties \citep[e.g., SN~2013ej;][]{fraser+14}.  We have shown that the dust shell around SN~2017eaw was compact ($4000~R_{\odot}$), and so it is likely that this dust was vaporized within the first few days after explosion.  This ought to be the case if the dust was mostly produced within the last decade before explosion, which is supported by the decrease in $F814W$ and enhancement in $4.5~\mu$m luminosity around $1000$~days before discovery.  Even if the light curves and spectra of SN~2017eaw (or any other SN~II-P) do not appear significantly reddened, it is still likely that circumstellar dust played a role in reddening the pre-explosion source.  This dust is more easily observed in mid-infrared bands, and so greater emphasis must be placed on obtaining imaging of SN progenitor stars beyond $2~\mu$m.

Once SN~2017eaw fades below the magnitude of this star, it will be straightforward to image the explosion site again and verify that the source has disappeared.  Deeper imaging can also be used to subtract any residual flux at the location of the progenitor and determine whether this system hosted a companion star.  The majority of massive stars will exchange mass with a companion at some point during their evolution \citep{sana+12,demink+14}.  Furthermore, there is direct evidence that SN~1993J left a surviving companion star \citep[e.g.,][]{maund+04,fox+14}, but there is no such evidence for SN~II-P progenitor stars.  Deep, targeted follow-up of nearby core-collapse SNe is an effective way to investigate the role of binary star evolution in producing these objects.

The proximity of SN~2017eaw and its location in a galaxy with $10$ luminous transients over the past hundred years presents a rare opportunity to study the environment and evolution of a SN~II in detail through very late phases.  Follow-up observations to study the metallicity, local environment, and any late-time circumstellar interaction around SN~2017eaw will help to resolve many lingering uncertainties in the properties of its progenitor system.
	
\smallskip\smallskip\smallskip\smallskip
\noindent {\bf ACKNOWLEDGMENTS}
\smallskip
\footnotesize

We would like to thank Ori Fox and Nathan Smith for helpful discussions.

This work is supported by NSF grant AST–1518052, the Gordon \& Betty Moore Foundation, the Heising-Simons Foundation, and by fellowships from the Alfred P.\ Sloan Foundation and the David and Lucile Packard Foundation to R.J.F.

This work is based in part on observations made with the {\it Spitzer Space Telescope}, which is operated by the Jet Propulsion Laboratory, California Institute of Technology, under a contract with NASA.  The {\it Hubble Space Telescope} (\hst) is operated by NASA/ESA. The \hst\ data used in this manuscript come from programs SNAP-9788, GO-14156, GO-14638, GO-14786, and SNAP-15166 (PIs Ho, Leroy, Long, Williams, and Filippenko, respectively). Some of our analysis is based on data obtained from the \hst\ archive operated by STScI. This work makes use of observations from the LCOGT network.

\textit{Facilities}: \hst\ (ACS/WFC3), \spitzer\ (IRAC), LCO (FLOYDS)

\bibliography{2017eaw}

\end{document}